\documentclass[a4paper,11pt]{article}
\pdfoutput=1
\usepackage[utf8x]{inputenc}
\usepackage{jheppub}

\usepackage[T1]{fontenc} 

\usepackage{amstext,amssymb}
\usepackage{amsmath}
\usepackage{graphicx}
\usepackage{color,soul}
\usepackage{xspace}
\usepackage{color}
\usepackage{units}
\usepackage{multirow}
\usepackage{slashed} 
\usepackage{comment}
\usepackage{subfigure}
\usepackage[]{hyperref}
\usepackage{appendix}
\usepackage{url}

\newcommand{\bea}{\begin{eqnarray}}
\newcommand{\eea}{\end{eqnarray}}
\newcommand{\nn}{\nonumber}

\title{\boldmath Neutrino mass, mixing and muon $g-2$ explanation in $U(1)_{L_\mu-L_\tau}$ extension of left-right theory}
\author[a]{\small \hspace*{-0.5cm} Chayan Majumdar}
\author[b]{\small , Sudhanwa Patra}
\author[a]{\small , Prativa Pritimita}
\author[a]{\small , Supriya Senapati}
\author[a]{\small , Urjit A. Yajnik}
\affiliation[a]{Department of Physics, Indian Institute of Technology Bombay, Powai, Mumbai-400076}
\affiliation[b]{Indian Institute of Technology Bhilai, GEC Campus, Sejbahar, Raipur-492015, India}
\emailAdd{chayan@phy.iitb.ac.in}
\emailAdd{sudhanwa@iitbhilai.ac.in}
\emailAdd{prativa@iitb.ac.in}
\emailAdd{supriya@phy.iitb.ac.in}
\emailAdd{yajnik@phy.iitb.ac.in}
\abstract{

We consider a gauged $U(1)_{L_\mu-L_\tau}$ extension of the left-right symmetric theory in order to simultaneously explain neutrino mass, mixing and 
the muon anomalous magnetic moment. We get sizeable contribution from the interaction of the new light gauge boson $Z_{\mu\tau}$ 
of the $U(1)_{L_\mu-L_\tau}$ symmetry with muons which can individually satisfy the current bounds on muon $(g-2)$ anomaly ($\Delta a_\mu$). The other positive contributions to $\Delta a_\mu$ come from the interactions of 
singly charged gauge bosons $W_L$, $W_R$ with heavy neutral fermions and that of neutral CP-even scalars with muons. The interaction of $W_L$ with heavy neutrino 
is facilitated by inverse seesaw mechanism which allows large light-heavy neutrino mixing and explains neutrino mass in our model. CP-even scalars with mass around few hundreds GeV can also satisfy the entire current muon anomaly bound. The results show that 
the model gives a small but non-negligible contribution to $\Delta a_\mu$ thereby eliminating the entire deviation in theoretical prediction and experimental result 
of muon $(g-2)$ anomaly. We have briefly presented a comparative study for symmetric and asymmetric left-right symmetric model in context of various contribution to $\Delta a_\mu$. We also discuss how the generation of neutrino mass is affected when left-right symmetry breaks down to Standard Model symmetry 
via various choices of scalars.}

\begin{document} 
\maketitle
\flushbottom
\section{Introduction}
\label{sec:intro}
While most of the theoretical predictions by Standard Model (SM) have been experimentally found to be correct to a very high precision, there lies a wide gap between  
SM's prediction of muon anomalous magnetic moment, $a_{\mu}= \frac{g_{\mu}-2}{2}$ and its measurement. The SM prediction can be summed up as
$a_{\mu}^{\text{SM}} = (11659183.0 \pm 4.8) \times 10^{-10}$ \cite{Tanabashi:2018oca, Blum:2013xva} whereas, the value obtained by 
Brookhaven National Laboratory (BNL) is $a_{\mu}^{\text{exp}} = (11659209.1 \pm 6.3)\times 10^{-10}$ \cite{Tanabashi:2018oca, Bennett:2006fi} with $\Delta a_{\mu}= (26.1 \pm 7.9)\times 10^{-10}$ \cite{Dev:2020drf} .
While a 3.3$\sigma$ deviation is achieved by BNL yet \cite{Bennett:2006fi}, a nearly 5$\sigma$ deviation is expected in the near future 
by Fermilab E989 \cite{Grange:2015fou} and of similar precision by J-PARC \cite{Abe:2019thb}. In principle the $a_{\mu}$ predicted by SM is a sum of contributions 
coming from QED, electroweak and hadronic sectors; 
\begin{align}
 a_{\mu}^{\text{SM}} = a_{\mu}^{\text{QED}} + a_{\mu}^{\text{electroweak}} + a_{\mu}^{\text{hadronic}} 
\end{align}
Among these three contributions, the theoretical uncertainty is believed to be coming from the hadronic loop contributions \cite{Davier:2017zfy,Keshavarzi:2018mgv,Davier:2019can} since the 
other two contributions have been verified with a high precision \cite{Gnendiger:2013pva,Aoyama:2017uqe}. A proposed experiment, namely MUonE \cite{Abbiendi:2016xup} aspires to reduce 
this theoretical uncertainty by determining the hadronic vacuum polarization more precisely. All these recent developments in the experimental muon sector 
surely ignites theoretical research that aim at eliminating or narrowing down this wide gap in the prediction and measurement.
Therefore recently many new physics scenarios have been explored in this context, for an incomplete list 
of which one may refer \cite{Jegerlehner:2009ry,Lindner:2016bgg,Ajaib:2014ana,Davoudiasl:2014kua,Rentala:2011mr,Kelso:2013zfa,Ky:2000ku,deS.Pires:2001da,Agrawal:2014ufa,Endo:2013lva,Ibe:2013oha,Everett:2001tq,Arnowitt:2003vw,Martin:2002eu,Taibi_2015,Altmannshofer:2016brv,Megias:2017dzd,Jana:2020pxx,Yamaguchi:2016oqz,Yin:2016shg,Endo:2019bcj,Dev:2017fdz}. 

Many of these new physics scenarios focus on $U(1)_{L_\mu-L_\tau}$ symmetry to address the anomaly because of the 
phenomenology associated with its gauge boson $Z_{\mu\tau}$.
The total lepton number, $L$, is a sum of individual lepton numbers $L_e$, $L_\mu$, $L_\tau$ and one can always choose the difference between any two 
individual lepton numbers like $L_e-L_\mu$, $L_\mu-L_\tau$, $L_e-L_\tau$ and gauge it to obtain an anomaly free theory. However, the gauged $U(1)_{L_\mu-L_\tau}$ 
symmetry is the most chosen one due to the fact that the parameters associated with $Z_{\mu \tau}$ gauge boson is not constrained by lepton and hadron colliders 
since it doesn't couple to electrons and quarks. Moreover, as per the constraints given by neutrino-trident experiments \cite{Altmannshofer:2014pba} a low mass of $\mathcal{O}$(100 MeV) can be allowed for this new gauge boson $Z_{\mu \tau}$ for a coupling as low as $g_{\mu \tau} \leq 10^{-3}$.

The $U(1)_{L_\mu-L_\tau}$ extension of SM has been extensively studied for explaining several issues like muon $(g-2)$ anomaly \cite{Garani:2019fpa, Heeck:2011wj}, 
dark matter \cite{Biswas:2016yjr}, orbital energy loss of a neutron star  \cite{Poddar:2019wvu} and so on. 
Several other works have explained how the associated $Z_{\mu\tau}$ gauge boson can ameliorate the tension in the late time and early time determination 
of Hubble constant \cite{Escudero:2019gzq}, unexpected dip in the energy spectrum of high energy cosmic neutrinos reported by the 
IceCube Collaboration \cite{Araki:2014ona} and also the deviations to neutrino oscillations due to long range forces \cite{Heeck:2010pg}.
Ref \cite{Dror:2019uea} says the vectors associated with a gauged $U(1)_{L_\mu-L_\tau}$ symmetry can induce an anomalously fast decay of the orbital period of 
neutron star binaries which might be used to discover any long-ranged muonic force associated with the binaries while ref \cite{Garani:2019fpa} 
explains how this gauge boson can possibly mediate interactions between dark matter particles and muons inside a neutron star. 
The effect induced by $L_\mu-L_\tau$ vector to enhanced production in neutrino decays, meson decays, neutrinoless double beta decays, and annihilations are discussed in ref \cite{Dror:2020fbh}.
The possible detection of this light $Z_{\mu\tau}$ boson is discussed in Ref \cite{Sirunyan:2018nnz,Araki:2017wyg,Baek:2001kca,Gninenko:2001hx}. 
However the $U(1)_{L_\mu-L_\tau}$ extension of SM can not accommodate neutrino mass untill and unless one adds a right-handed neutrino to the model. 
Such attempts have been made in ref \cite{Biswas:2016yan, Biswas:2016yjr} , where the authors explain neutrino mass by adding three 
right-handed neutrinos to the model.  In ref \cite{Ma:2001md} neutrino masses with bimaximal mixing is obtained just by adding one right-handed neutrino 
to the extended SM framework. A similar framework \cite{Choubey:2004hn} also predicts quasi-degenerate neutrino masses. On the other hand, the left-right symmetric model (LRSM) \cite{Mohapatra:1974gc, Pati:1974yy, Senjanovic:1975rk,Senjanovic:1978ev,Mohapatra:1979ia,Mohapatra:1980yp,Pati:1973uk,Pati:1974vw} is a SM extension which clearly gives us a unified answer to small neutrino mass generation as well as parity violation problem in low-energy weak interactions. LRSM naturally hosts a right-handed neutrino and offers wider possibilities of explaining neutrino mass, lepton number violation, lepton flavour violation with rich phenomenology at low scale. In particular we shall see that an interplay between the right handed gauge bosons and the mass mechanisms for the neutrinos makes an important contribution.

Thus with the motivation of explaining neutrino mass, mixing and muon $(g-2)$ anomaly in a single framework we reach for the LRSM and augment it with the $U(1)_{L_\mu-L_\tau}$ symmetry. In manifest LRSM neutrino mass can be explained by canonical 
seesaw mechanism, but it cannot be verified by collider experiments since a very high right-handed breaking scale $(10^{14}~\text{GeV})$ is associated with the mechanism. Thus in general 
extra particles are added to LRSM in order to generate neutrino mass by various low-scale seesaw mechanisms like linear seesaw, inverse seesaw \cite{Mohapatra:1986bd, Akhmedov:1995vm, Akhmedov:1995ip, Barr:2003nn, Barr:2005ss,Nomura:2019xsb,Sahu:2020tqe,Sruthilaya:2017mzt,Deppisch:2015cua,Humbert:2015yva,Parida:2012sq}, double seesaw
etc \cite{Tello:2010am,Barry:2013xxa,Dev:2013vxa,Nemevsek:2011hz,Dev:2014iva,Das:2012ii,Bertolini:2014sua,Dhuria:2015cfa,Borah:2013lva,Chakrabortty:2012mh,Deppisch:2015cua, 
Majumdar:2018eqz, Bambhaniya:2015ipg, Dev:2014xea}. In particular, we take interest in inverse seesaw in our extended LRSM to explain neutrino mass which also allows large light-heavy neutrino mixing 
and thus leads to sizeable contributions to the muon anomalous magnetic moment via left-handed singly-charged SM gauge boson interaction with heavy neutrino. Apart from the usual fermions and scalars present in a manifest LRSM, the model 
contains three sets of extra sterile fermions and one extra scalar. While the extra sterile fermions help in creating the plot for  inverse seesaw, the extra scalar helps in breaking the $U(1)_{L_\mu-L_\tau}$ symmetry and also in implementing the inverse seesaw in the model.
The $Z_{\mu\tau}$ boson originated from the breaking of $U(1)_{L_\mu-L_\tau}$ symmetry helps in ameliorating $\Delta a_\mu$ when it gets mass around $150~ \text{MeV}$. 
Moreover our predictions on the mass of $Z_{\mu\tau}$ and its coupling $g_{\mu\tau}$ lie well below the constraint given by ref \cite{DiFranzo:2015qea}. We also discuss various symmetry breaking chains from LRSM to 
SM with different choices of scalars to see how it affects the generation of neutrino mass. Also we have shown that lighter neutral CP-even scalars can also satisfy the current as well as 1$\sigma$ bound on muon anomaly individually if they possess mass around $0.5-2$ TeV.

The rest of the paper is organised as follows. In Sec \ref{sec:model} we present the particle content of the extended LRSM and discuss the symmetry breaking of 
$U(1)_{L_\mu-L_\tau}$ symmetry and left-right symmetry down to low energy theory. We also discuss two different scenarios of neutrino mass generation with the help 
of doublet scalars in \ref{doublets} and triplet scalars in \ref{deltaLR}. In Sec \ref{sec:numass} we discuss the generation of neutrino mass and mixing via extended inverse seesaw mechanism. 
In Sec \ref{sec:muon_anomaly_prediction} we analytically study the new contributions to $\Delta a_\mu$ arising from different vector bosons and scalars present in the model.
In Sec \ref{sec:results} we estimate the contributions numerically and present the results. This section also contains several plots of $\Delta a_\mu$ $vs$ mass of 
mediators to check the sensitivity of our theoretical results to experimental bounds. In Sec \ref{sec:conclusion} we summarize and conclude the work.

\section{The Model}
\label{sec:model}
The model is an extension of manifest left-right theory with additional 
$U(1)$ gauge symmetry where the difference between muon and tau lepton numbers is gauged. 
The model is governed by the gauge group, 
\begin{align}
\mathbb{G^{\mu\tau}_{LR}} \equiv  SU(2)_L\times SU(2)_R\times U(1)_{B-L} \times SU(3)_C \times U(1)_{L_{\mu}-{L_\tau}}
\end{align}

Within manifest LRSM which is based on the gauge group 
$SU(2)_L\times SU(2)_R\times U(1)_{B-L}\times SU(3)_C$ and consists of usual quarks ($q_{L,R}$), leptons ($\ell_{L,R}$), Higgs bidoublet $\Phi$ and 
triplets $\Delta_{L,R}$ (presented in Table \ref{tab:LRSM}) the light neutrino masses can be generated by type-I+II seesaw mechanism \cite{Barry:2013xxa,Chakrabortty:2012mh,Dev:2013vxa,Das:2012ii,Bertolini:2014sua,Borah:2016iqd,Borah:2015ufa}, 
$$m_\nu= - M_D M^{-1}_R M^T_D + M_L = m^{I}_\nu + m^{II}_\nu\,,$$
\begin{table}[h]
\begin{center}
\begin{tabular}{|c||c|c|c|c|c|}
 \hline
          & Fields      & $ SU(2)_L$ & $SU(2)_R$ & $B-L$ & $SU(3)_C$ \\
\hline
 Fermions &$q_L$     &  2         & 1         & 1/3   & 3   \\
 & $q_R$     &  1         & 2         & 1/3   & 3   \\
 & $\ell_L$  &  2         & 1         & -1    & 1   \\
 & $\ell_R$  &  1         & 2         & -1    & 1   \\
 \hline
 Scalars & $\Phi$   &  2         & 2         &  0     & 1   \\
 & $\Delta_L$    &  3         & 1         & 2     & 1   \\
 & $\Delta_R$    &  1         & 3         & 2     & 1   \\
\hline
\end{tabular}
\end{center}
\caption{Particle content of the manifest left-right symmetric theories. }
\label{tab:LRSM}
\end{table}
where $M_L (M_R)$ represents the Majorana mass term for light left-handed (heavy right-handed) Majorana neutrinos 
arising from respective VEVs of left-handed (right-handed) scalar triplet and $M_D$ is the Dirac neutrino 
mass matrix connecting light-heavy neutrinos. Here, the scale of right-handed neutrino mass ($M_R$) is related to the non-zero VEV of right-handed scalar triplet which is responsible for spontaneous symmetry breaking of LRSM to SM.
The sub-eV scale of light neutrino 
mass, as hinted by oscillation experiments, is connected to a very heavy right-handed scale i.e, $10^{15}~$GeV (in generic scenarios) clearly making it inaccessible to current and planned accelerator experiments. On the other hand, when LRSM breaks around TeV scale, the gauge bosons $W_R$, $Z_R$, right-handed neutrinos $N_R$ and scalar triplets $\Delta_{L,R}$ get TeV scale mass  that allows several lepton number violating signatures at LHC as well as low energy experiments like neutrinoless double beta decay. 
The left-right mixing (or light-heavy neutrino mixing), which depends on Dirac neutrino mass $M_D$, plays an important role in giving large new contribution to neutrinoless double beta decay, other LNV signatures at colliders as well as LFV processes.
This gives the motivation to explore alternative class of left-right symmetric model with large value of $M_D$ and thereby large light-heavy neutrino mixing which can contribute positively to $\Delta a_\mu$.

A number of LRSM variants have been explored in literature \cite{Majumdar:2018eqz,FileviezPerez:2016erl,PhysRevLett.60.1813,PhysRevLett.62.1079,Gu:2010zv,Borah:2016hqn,Bolton:2019bou} where spontaneous symmetry breaking is implemented with scalar bidoublet having $B-L=0$ and 
Higgs doublets having $B-L=1$ which leads to neutrino mass being generated by either simple Dirac mass terms or low scale seesaw mechanisms like 
inverse seesaw, linear seesaw etc. 
In this model, for the generation of neutrino mass we take interest in inverse seesaw mechanism since it allows 
large light-heavy neutrino mixing and this mixing facilitates the interaction of singly charged vector boson with heavy neutrinos 
which contributes positively to $\Delta a_\mu$. Before we move on to the working of inverse seesaw mechanism in the considered model, let's have a clear picture of how the generation of neutrino mass is affected within 
various symmetry breaking of LRSM-SM chains.

At first, the spontaneous symmetry breaking (SSB) of $\mathbb{G^{\mu\tau}_{LR}}$ down to left-right theory $\mathbb{G_{LR}}$ is achieved 
by assigning a non-zero VEV to a scalar $\chi$ which is singlet under left-right symmetry but non-trivially charged under $U(1)_{L_{\mu}-{L_\tau}}$. 
Further, the SSB of LRSM to SM can happen in the following three ways;
\begin{itemize}
\item with Higgs doublets $H_L \oplus H_R$, 
\item with Higgs triplets $\Delta_L \oplus \Delta_R$,
\item with the combination of doublets and triplets $H_L \oplus H_R$ and $\Delta_L \oplus \Delta_R$. 
\end{itemize}
Now, as usual the SSB of SM to low energy theory occurs when the scalar bidoublet $\Phi$ takes non-zero vev and  that generates masses for charged leptons and quarks. 

\begin{table}[h]
\begin{center}
\begin{tabular}{|c||c|c|c|c|c|c|}
 \hline
  & Fields     & $SU(2)_L$ & $SU(2)_R$ & $U(1)_{B-L}$ & $SU(3)_C$ & $U(1)_{L_\mu-L_\tau}$\\
\hline
 Fermions & $\ell_{e_L}$  &  2         & 1         & -1    & 1  & 0 \\
 & $\ell_{\mu_L}$  &  2         & 1         & -1    & 1  & 1 \\
 & $\ell_{\tau_L}$  &  2         & 1         & -1    & 1  & -1 \\
 & $\ell_{e_R}$  &  1         & 2         & -1    & 1  & 0 \\
 & $\ell_{\mu_R}$  &  1         & 2         & -1    & 1  & 1 \\
 & $\ell_{\tau_R}$  &  1         & 2        & -1    & 1  & -1 \\ 
 \hline
 Scalars & $\Phi$   &  2         & 2         &  0     & 1  & 0 \\
  & $H_L$    &  2         & 1         & 1     & 1   & 0 \\
  & $H_R$    &  1         & 2         & 1     & 1   & 0 \\
  & $\chi$    &  1        & 1         & 0     & 1  & 1 \text{or} 2 \\
\hline
\end{tabular}
\end{center}
\caption{Particle content of left-right theories extended with 
$U(1)_{L_{\mu}-{L_\tau}}$ gauge symmetry where fermion sector is limited to leptons and scalar sector contains the bidoublet $\Phi$, doublets $H_{L,R}$ and a singlet $\chi$.}
\label{tab:mutau_LRSM_HLR}
\end{table}

\subsection{Neutrino Masses with LRSM-SM symmetry breaking via $H_R, H_L$}
\label{doublets}
In this minimal version, $H_R$ breaks the left-right symmetry to SM while $H_L$ is required for left-right invariance. The scalar bidoublet $\Phi$ is required for SM symmetry breaking to low energy theory and $\chi$ is needed for the spontaneous symmetry breaking of $\mathbb{G^{\mu\tau}_{LR}}$ down to left-right theory the spontaneous symmetry breaking (SSB) of $\mathbb{G^{\mu\tau}_{LR}}$ down to left-right theory $\mathbb{G_{LR}}$ as mentioned earlier.
The leptons and scalars are displayed in Table \ref{tab:mutau_LRSM_HLR}. 
The allowed Yukawa interactions for leptons are given by,
\begin{eqnarray}
\hspace*{-0.4cm}
-\mathcal{L}_{Yuk} &\supset& \,\overline{\ell_{e_L}} \left[Y_\ell \Phi + \tilde{Y}_\ell \widetilde{\Phi} \right] \ell_{e_R}
+\,\overline{\ell_{\mu_L}} \left[Y_\ell \Phi + \tilde{Y}_\ell \widetilde{\Phi} \right] \ell_{\mu_R}+
\,\overline{\ell_{\tau_L}} \left[Y_\ell \Phi + \tilde{Y}_\ell \widetilde{\Phi} \right] \ell_{\tau_R}+\mbox{h.c.}
\label{eqn:LR-Yuk}
\end{eqnarray}
with $\tilde{\Phi} = \tau_2 \Phi^{\ast} \tau_2$.

The vev structure for the Higgs spectrum can be depicted as follows:
\begin{center}
$\langle H_R \rangle = \begin{pmatrix}
0 \\
v_R
\end{pmatrix}$,~~
$\langle H_L \rangle = \begin{pmatrix}
0 \\
v_L
\end{pmatrix}$,~~
$\langle \Phi \rangle = \begin{pmatrix}
v_1 & 0 \\
0 & v_2
\end{pmatrix}$\,.
\end{center}
After SSB, the charged fermion as well as light neutrino mass matrices are found to be diagonal in structure due to presence of $U(1)_{L_{\mu}-{L_\tau}}$ gauge symmetry. This is the important prediction of the model giving simplified relation for PMNS mixing matrix as $U_{\rm PMNS} \equiv U_{\nu}$.

The non-zero masses for light neutrinos (which are Dirac fermions) can be explained by adjusting Yukawa couplings through 
the non-zero VEVs of scalar bidoublet. From the Yukawa interactions given in Eq.(\ref{eqn:LR-Yuk}); with $Y_\ell \ll \tilde{Y}_\ell$, $v_2 \ll v_1$, 
the masses for charged leptons and the light neutrinos can be expressed as,
\begin{eqnarray}
&&  M_\ell \simeq \tilde{Y}_\ell v^*_1\,, \quad \quad M_D^\nu \simeq v_1\left(Y_\ell+M_\ell \frac{v_2}{v^2_1} \right)\,.
\end{eqnarray}
Even though this framework holds a minimal 
(in terms of $SU(2)$ representation) scalar spectrum it can not provide Majorana mass for neutrinos and thus forbids any signature of lepton number violation.

\subsection{Neutrino Masses with LRSM-SM symmetry breaking via $\Delta_R, \Delta_L$}
\label{deltaLR}
In Table \ref{tab:mutau_LRSM_HLR} if we replace the doublets $H_L$, $H_R$ by triplets $\Delta_L$, $\Delta_R$ then the model offers a better possibility from phenomenology point of view since in this case 
 Majorana masses can be generated for light and heavy neutrinos. If the symmetry breaking occurs at few TeV scale, these Majorana neutrinos can mediate neutrinoless double beta decay process whose observation would confirm lepton number violation in nature. Lepton number violation 
 can also be probed via smoking-gun same-sign dilepton signatures at collider experiments.
 The interaction terms involving scalar triplets and leptons in the left-right theories with extra $U(1)$ symmetry are given by
\begin{eqnarray}
-\mathcal{L}_{Yuk} &\supset& 
\,\overline{\ell_{e_L}} \left[Y_\ell \Phi + \tilde{Y}_\ell \widetilde{\Phi} \right] \ell_{e_R}
+\,\overline{\ell_{\mu_L}} \left[Y_\ell \Phi + \tilde{Y}_\ell \widetilde{\Phi} \right] \ell_{\mu_R}+
\,\overline{\ell_{\tau_L}} \left[Y_\ell \Phi + \tilde{Y}_\ell \widetilde{\Phi} \right] \ell_{\tau_R} \nonumber \\
&+&
\bigg[f_{ee} \overline{(\ell_{e_L})^c} \ell_{e_L} 
+f_{\mu \tau} \overline{(\ell_{\mu_L})^c} \ell_{\tau_L}
+f_{\tau \mu} \overline{(\ell_{\tau_L})^c} \ell_{\mu_L}
\bigg] \Delta_L 
\nonumber \\
&+&
\bigg[f_{ee} \overline{(\ell_{e_R})^c} \ell_{e_R} 
+f_{\mu \tau} \overline{(\ell_{\mu_R})^c} \ell_{\tau_R}
+f_{\tau \mu} \overline{(\ell_{\tau_R})^c} \ell_{\mu_R}
\bigg] \Delta_R 
+ \mbox{h.c.}
\label{eqn:LR-Yuk-triplets}
\end{eqnarray}
with the corresponding vevs
\begin{center}
$\langle \Delta_L \rangle = \begin{pmatrix}
0 & 0 \\
v_L & 0
\end{pmatrix},~~
\langle \Delta_R \rangle = \begin{pmatrix}
0 & 0 \\
v_R & 0
\end{pmatrix},~~
\langle \Phi \rangle = \begin{pmatrix}
v_1 & 0 \\
0 & v_2
\end{pmatrix}$
\end{center}
Using Eq.(\ref{eqn:LR-Yuk-triplets}), the structure of the masses 
for neutral leptons in the basis $\left(\nu_L, N^c_R\right)$ can be written as,
\begin{equation}
\mathbb{M}= \left( \begin{array}{cc}
              M_{L} & M_{D}   \\
              M^T_{D} & M_{R}
                      \end{array} \right) \, ,
\label{eqn:numatrix}       
\end{equation}
where, $M_D$ represents Dirac neutrino mass matrix, $M_L (M_R)$ denotes Majorana mass matrix arising from the non-zero vev of LH (RH) scalar triplet. 
The mass matrices $M_D$, $M_L$ and $M_R$ can be written explicitly as follows (considering $f_{\mu \tau} = f_{\tau \mu}$ and $f_{\mu \mu}=f_{\tau \tau}$ for sake of simplicity),
\begin{eqnarray}
&&M_D = \begin{pmatrix}
Y_{11}v_2 +\tilde{Y}_{11} v_1 & 0 & 0\\
0 & Y_{22}v_2 +\tilde{Y}_{22} v_1 & 0\\
0 & 0 & Y_{33} v_2 +\tilde{Y}_{33} v_1\\
\end{pmatrix} = \begin{pmatrix}
a & 0 & 0\\
0 & b & 0\\
0 & 0 & c\\
\end{pmatrix}\, , \nonumber \\
&&M_{L,R} = \begin{pmatrix}
f_{ee} & 0 & 0\\
0 & 0 & f_{\mu \tau}\\
0 & f_{  \mu \tau} & 0\\
\end{pmatrix}\, \frac{v_{L,R}}{\sqrt{2}} , \nonumber \\
\end{eqnarray}

Now using seesaw approximation $M_R \gg M_D$ and $M_L \to 0$, the light neutrino mass can be generated via type-I seesaw formula as shown below, 
\begin{eqnarray}
&&m_{\nu}^I = -M_{D}\, M_{R}^{-1}\, M_{D}^{T} \nonumber \\
&=&\begin{pmatrix}
a & 0 & 0\\
0 & b & 0\\
0 & 0 & c\\
\end{pmatrix} \cdot 
\begin{pmatrix}
f_{ee} \frac{v_R}{\sqrt{2}} & 0 & 0\\
0 & 0 & f_{\mu \tau} \frac{v_R}{\sqrt{2}}\\
0 & f_{  \mu \tau} \frac{v_R}{\sqrt{2}} & 0\\
\end{pmatrix}^{-1} \cdot 
\begin{pmatrix}
a & 0 & 0\\
0 & b & 0\\
0 & 0 & c\\
\end{pmatrix}^T  \nonumber \\
&=&\begin{pmatrix}
\frac{\sqrt{2}a^2}{f_{ee}v_R} & 0 & 0 \\
0 & 0 & \frac{\sqrt{2}bc}{f_{\mu \tau} v_R} \\
0 & \frac{\sqrt{2}bc}{f_{\mu \tau} v_R} & 0
\end{pmatrix} 
\end{eqnarray}

From this light neutrino mass matrix $m_\nu^I$, the corresponding mass eigenvalues for light neutrino mass eigenstates can be obtained which are,
$\lbrace -\frac{\sqrt{2}bc}{f_{\mu \tau} v_R}, \frac{\sqrt{2}bc}{f_{\mu \tau} v_R}, \frac{\sqrt{2}a^2}{f_{ee} v_R} \rbrace$. 
However, two mass eigenstates with eigenvalues $\frac{\sqrt{2}bc}{f_{\mu \tau} v_R}$ (ignoring the relative negative sign) are degenerate here, which implies 
either the solar neutrino mass difference ($\Delta m^2_{\text{sol}}$ ) or the atmospheric neutrino mass difference ($\Delta m^2_{\text{atm}}$) vanishes. This is in disagreement with the neutrino experimental data at 3$\sigma$ interval of global fit by NuFIT 4.1~\cite{Esteban:2018azc}, 
\begin{align}
&{\rm NO}: \Delta m^2_{\rm atm}=[2.432, 2.618]\times 10^{-3}\ {\rm eV}^2,\
\Delta m^2_{\rm sol}=[6.79, 8.01]\times 10^{-5}\ {\rm eV}^2,\\
&\sin^2\theta_{13}=[0.02046, 0.02440],\ 
\sin^2\theta_{23}=[0.427, 0.609],\ 
\sin^2\theta_{12}=[0.275, 0.350],\nn\\
&{\rm IO}: \Delta m^2_{\rm atm}=[2.416, 2.603]\times 10^{-3}\ {\rm eV}^2,\
\Delta m^2_{\rm sol}=[6.79, 8.01]\times 10^{-5}\ {\rm eV}^2,\\
&\sin^2\theta_{13}=[0.02066, 0.02461],\ 
\sin^2\theta_{23}=[0.430, 0.612],\ 
\sin^2\theta_{12}=[0.275, 0.350].\nn
\end{align}

This degeneracy can be wiped out by introducing another pair of triplet scalars 
$\Delta_L^{\prime} \oplus \Delta_R^{\prime}$ with $L_{\mu} - L_{\tau} = 2$. Now we can write additional Yukawa terms 
allowed by the $U(1)_{L_{\mu} - L_{\tau}}$ symmetry as,
\begin{align}
-\mathcal{L}_{Yuk}^{\text{new}} \supset f_{\mu\mu} (\ell^T_{\mu R} \Delta^{\prime \dagger}_R \ell_{\mu R} + \ell^T_{\tau R} \Delta^{\prime}_R \ell_{\tau R}) + R \leftrightarrow L   
\end{align}
 With these new permissible terms in the Yukawa sector, we can write the corresponding $M^{\prime}_{L,R}$ matrix as,
 \begin{equation}
M^{\prime}_{L,R} = \begin{pmatrix}
f_{ee}\frac{v_{L,R}}{\sqrt{2}} & 0 & 0\\
0 & f_{\mu\mu}\frac{v^{\prime}_{L,R}}{\sqrt{2}} ~&~ f_{\mu \tau}\frac{v_{L,R}}{\sqrt{2}}\\
0 & f_{  \mu \tau}\frac{v_{L,R}}{\sqrt{2}} ~&~ f_{\mu\mu}\frac{v^{\prime}_{L,R}}{\sqrt{2}}\\
\end{pmatrix} 
 \end{equation} 
where $v^{\prime}_{L,R} = \langle \Delta^{\prime}_{L,R} \rangle$. Now using the seesaw approximation $M_{R}^{\prime} \gg M_D$ 
and $M_{L}^{\prime} \rightarrow 0$, the light neutrino mass matrix can be expressed via type-I seesaw formula as,
\begin{align}
m_{\nu}^{\prime I} &= -M_D M_{R}^{\prime -1} M_D^{T} \nonumber \\
&= \begin{pmatrix}
\frac{\sqrt{2}a^2}{f_{ee}v_R} & 0 & 0 \\
0 & \frac{\sqrt{2}b^2f_{\mu\mu}v^{\prime}_R}{-f_{\mu\tau}^2 v_{R}^2+f_{\mu\mu}^2 v_{R}^{\prime 2}} ~&~ \frac{\sqrt{2}bc f_{\mu\tau}v_R}{f_{\mu\tau}^2 v_{R}^2-f_{\mu\mu}^2 v_{R}^{\prime 2}} \\
0 & \frac{\sqrt{2}bc f_{\mu\tau}v_R}{f_{\mu\tau}^2 v_{R}^2-f_{\mu\mu}^2 v_{R}^{\prime 2}} ~&~ \frac{\sqrt{2}c^2f_{\mu\mu}v^{\prime}_R}{-f_{\mu\tau}^2 v_{R}^2+f_{\mu\mu}^2 v_{R}^{\prime 2}}
\end{pmatrix}
\end{align} 
Here all three mass eigenvalues are non-degenerate which can be represented as, $\lbrace \frac{\sqrt{2}a^2}{f_{ee}v_R}, m_{\nu}^{I a} \pm m_{\nu}^{I b} \rbrace$ with 
\begin{center}
$m_{\nu}^{I a} = \frac{-b^2 f_{\mu\mu}v^{\prime}_R-c^2 f_{\mu\mu}v^{\prime}_R}{\sqrt{2}(f_{\mu\tau}^2v_{R}^2-f_{\mu\mu}^2v_{R}^{\prime 2})}$, \\
 $m_{\nu}^{I b} = \frac{\sqrt{4b^2 c^2 f_{\mu\tau}^2v_{R}^2 + b^4 f_{\mu\mu}^2 v_{R}^2 -2b^2 c^2 f_{\mu\mu}^2 v_{R}^{\prime 2} + c^4 f_{\mu\mu}^2 v_{R}^{\prime 2}}}{\sqrt{2}(f_{\mu\tau}^2v_{R}^2-f_{\mu\mu}^2v_{R}^{\prime 2})}$
 \end{center}
Though the introduction of two extra scalar triplets $\Delta^{\prime}_L \oplus \Delta^{\prime}_R$ 
 saves us from apparent inconsistency in the explanation of current-day neutrino oscillation data, the particle content of the model becomes crowded and it no more remains minimal.
  
\section{LRSM Inverse Seesaw (LISS) for Neutrino Masses}
\label{sec:numass}

We have already discussed in previous section that the sub-eV scale  neutrino masses can be generated either by canonical see-saw mechanism which requires a very high ($> 10^{14}$~GeV) seesaw scale and therefore cannot be verified by present-day colliders
or with very much suppressed value of Dirac neutrino Yukawa coupling. As an alternative,
we explain one of the low scale seesaw mechanism, i.e, LRSM inverse seesaw (LISS)~\cite{Nandi:1985uh,Mohapatra:1986aw,Mohapatra:1986bd,Dev:2009aw,Sahu:2004sb,Awasthi:2013ff,Pritimita:2016fgr,Majee:2008mn,Kang:2006sn,Hewett:1988xc,Blanchet:2010kw,Dias:2012xp,Das:2017kkm} in our model where the left-right symmetry breaking occurs at few TeV. This symmetry breaking generates TeV scale masses for $W_R$, $Z_R$ gauge bosons which fall within the LHC range and the inverse seesaw mechanism provides large light-heavy neutrino mixing. 
As a result the large mixing between sub-TeV scale heavy neutrinos with sub-eV scale light neutrinos, within left-right inverse seesaw scheme, offers, 
\begin{itemize}
 \item sizeable contribution to muon $g-2$ anomaly arising form purely left-handed currents with the exchange of sub-TeV masses for sterile neutrinos in LISS scheme, 
 \item dominant contribution to lepton flavour violating (LFV) decays, non-unitarity effects in leptonic sector,
 \item interesting collider signatures verifiable at LHC.
\end{itemize}
For implementing LISS, we consider an extra sterile neutrino $S_L$ per generation along with the usual leptons, scalars (bidoublet $\Phi$, doublets $H_{L,R}$ and $\chi$) presented in Table~\ref{tab:mutau_LRSM_HLR}. The relevant Yukawa interaction Lagrangian for LISS invariant under $U(1)_{L_\mu-L_\tau}$ symmetry is given as sum of different components,
\begin{align}\label{Lmutaulag}
 -\mathcal{L}_{\text{LISS}}=&\mathcal{L}_{\nu_L N_R}+\mathcal{L}_{N_R S_L}+\mathcal{L}_{S_L S_L}\,,
\end{align}
where the individual components are given as follows:

\noindent
\underline{\bf Generic Dirac Neutrino Mass Matrix, $\mathcal{L}_{\nu_L N_R}$:}\\
The usual Dirac Yukawa interaction Lagrangian that allows Dirac mass terms for charged leptons and neutrinos) consistent with the $U(1)_{L_{\mu}-L_{\tau}}$ gauge symmetry is given by,
\begin{align}
\mathcal{L}_{\nu_L N_R}& \supset \overline{\ell_L}(Y \Phi + \tilde{Y} \tilde{\Phi}) \ell_R \nonumber \\
& =\overline{\ell_{e_L}} [M_i]^{ee} \ell_{e_R}+\overline{\ell_{\mu_L}}[M_i]^{\mu\mu} \ell_{\mu_R}+\overline{\ell_{\tau_L}}[M_i]^{\tau\tau} \ell_{\tau_R}
\label{LnuN}
\end{align}
where, $M_i=M_\ell, M^\nu_D \equiv M_D$ are the corresponding Dirac mass matrices for charged leptons and neutrinos respectively. The imposition of extra $U(1)_{L_\mu-L_\tau}$ symmetry to the left-right theories results in diagonal Dirac mass matrices for charged leptons and neutrinos as,
\begin{eqnarray}
&& M_\ell = \begin{pmatrix}
Y_{11}v_1+\tilde{Y}_{11}v_2 & 0 & 0\\
0 & Y_{22}v_1+\tilde{Y}_{22}v_2 & 0\\
0 & 0 & Y_{33}v_1+\tilde{Y}_{33}v_2\\
\end{pmatrix} \nonumber \\
&& M_D = \begin{pmatrix}
Y_{11}v_2+\tilde{Y}_{11}v_1 & 0 & 0\\
0 & Y_{22}v_2+\tilde{Y}_{22}v_1 & 0\\
0 & 0 & Y_{33}v_2+\tilde{Y}_{33}v_1\\
\end{pmatrix} = \begin{pmatrix}
a & 0 & 0 \\
0 & b & 0 \\
0 & 0 & c
\end{pmatrix}\,.
\end{eqnarray}

\noindent
\underline{\bf Dirac Mass term between $N_R$ and $S_L$, $\mathcal{L}_{N_R S_L}$:}\\
The corresponding Yukawa term gives rise to the mixing matrix $M$ between $N_R$ and $S_L$ as,
\begin{equation}
\mathcal{L}_{N_R S_L} \supset Y_{RS}\overline{\ell} \tilde{H}_R S_L = Y_{RS} \langle \tilde{H}_R \rangle \left[\overline{\ell_{e_R}} S_{e_L}+\overline{\ell_{\mu_R}}S_{\mu_L}+\overline{\ell_{\tau_R}} S_{\tau_L}\right]
\label{LNS}
\end{equation}
The corresponding mixing matrix is also found to be diagonal as,
\begin{center}
$M = \begin{pmatrix}
M_{11} & 0 & 0 \\
0 & M_{22} & 0 \\
0 & 0 & M_{33}\\
\end{pmatrix}
$
\end{center} 
whose diagonal entries are proportional to $\langle \tilde{H}_R \rangle = v_R$.

\noindent
\underline{\bf Bare Majorana Mass term for $S_L$, $\mathcal{L}_{S_L S_L}$:}\\
Now, we focus on the generation of bare Majorana mass term for sterile neutrinos and the $U(1)_{L_{\mu}-L_{\tau}}$ gauge group allows the terms involving 
extra sterile neutrinos as, 
\begin{eqnarray}
\mathcal{L}_{S_L S_L} &=& \mu S^T_L S_L \nonumber \\
&=&\bigg[\mu_{ee} S_{e_L}^T S_{e_L} +\mu_{\mu \tau} S_{\mu_L}^T S_{\tau_L} + \mu_{\mu\tau} S_{\tau_L}^T S_{\mu_L} \bigg]
\label{LSS}
\end{eqnarray}
So the bare Majorana mass matrix structure for extra sterile neutrinos can be expressed as,
\begin{equation}
\mu = \begin{pmatrix}
\mu_{ee} & 0 & 0 \\
0 & 0 & \mu_{\mu\tau} \\
0 & \mu_{\mu\tau} & 0
\end{pmatrix}
\end{equation}

Thus, the complete $9 \times 9$ neutral fermion mass matrix in the basis of 
$(\nu_L, N_R, S_L)$ is read as,
\begin{equation}
\mathbb{M} = \begin{pmatrix}
0 & M_D & 0 \\
M_D^T & 0 & M^T \\
0 & M & \mu \\
\end{pmatrix}
\label{whole}
\end{equation}
Using eq.~\ref{whole} with mass hierarchy $M > M_D \gg \mu$, we can write the expression for
Majorana mass ($m_{\nu}$) for light neutrinos and pseudo-Dirac mass term ($m_H$) for heavy neutrinos in LISS as~\cite{Mohapatra:1986aw,Dev:2009aw,Sahu:2004sb,Awasthi:2013ff,Pritimita:2016fgr,Majee:2008mn,Kang:2006sn,Hewett:1988xc,Blanchet:2010kw,Dias:2012xp,Das:2017kkm,Dev:2017fdz},
\begin{equation}
m_{\nu} = \left(\frac{M_D}{M}\right) \mu \left(\frac{M_D}{M}\right)^T 
\label{eq:light}
\end{equation}
\begin{equation}
m_H = -(\pm M - \mu /2)
\end{equation} 

The beautiful aspect of the low scale inverse seesaw scheme is that it allows sub-eV scale of light neutrinos with large value of $M_D$ and $M$ as,
$$\left( \frac{m_\nu}{\mbox{0.1\, eV}}\right) = \left(\frac{M_D}{\mbox{100\, GeV}} \right)^2 
 \left(\frac{\mu}{\mbox{keV}}\right) \left(\frac{M}{10^4\, \mbox{GeV}} \right)^{-2}\,.$$

Even with $M (\sim $ sub TeV scale), we can have sizeable light-heavy neutrino 
mixing ($M_D/M \simeq \mathcal{O}(0.1-1)$) which can give rise to large charged LFV decay channels as $\mu \rightarrow e\gamma, \tau \rightarrow \mu$ and $0\nu\beta\beta$ effects~\cite{Awasthi:2013ff}. Now from eq.(\ref{eq:light}), in this LRSM inverse seesaw approximation, we can express the light neutrino mass matrix as,
\begin{equation}
m_{\nu}^{\text{LISS}} = \begin{pmatrix}
\frac{a^2 \mu_{ee}}{M_{11}^2} & 0 & 0 \\
0 & 0 & \frac{bc \mu_{\mu\tau}}{M_{22}M_{33}} \\
0 & \frac{bc \mu_{\mu\tau}}{M_{22}M_{33}} & 0 
\end{pmatrix} 
\end{equation}
which delivers light neutrino mass eigenstates with degenerate eigenvalues $\lbrace \frac{a^2 \mu_{ee}}{M_{11}^2} , -\frac{bc \mu_{\mu\tau}}{M_{22}M_{33}}, 
\frac{bc \mu_{\mu\tau}}{M_{22}M_{33}} \rbrace$ similar to the previous situation given in eq.(\ref{deltaLR}). Since the mass matrices $M_D$ and $M$ are diagonal in structure, the non-degenerate light neutrino masses consistent with observed values $\Delta m^{2}_{\rm sol}$ and $\Delta m^{2}_{\rm atm}$ can be achieved by suitable modification in the $\mu$ matrix. The modification in the matrix structure of $\mu$ matrix can be implemented with the inclusion of extra terms in the $\mu$ matrix which may be either of off-diagonal or diagonal in nature. Therefore, the extra singlet scalar $\chi$ with non-zero $U(1)_{L_{\mu} - L_{\tau}}$ charge which was originally introduced for spontaneous symmetry breaking of $U(1)_{L_\mu-L_\tau}$ symmetry can remove this degeneracy without affecting the usual left-right symmetry. We call this scenario as `Extended LRSM with Inverse Seesaw (ELISS)'. The introduction of $\chi$ allows additional Yukawa-like terms in the Lagrangian and now the total Lagrangian for ELISS scenario becomes, 
\begin{eqnarray}
&&\mathcal{L}_{\text{ELISS}} = \mathcal{L}_{\text{LISS}} +  \mathcal{L}_{\chi}
\end{eqnarray}
where $\mathcal{L}_\chi$ is the correction terms to the LISS lagrangian due to the introduction of new scalar $\chi$. \\
 \underline{\bf $\mathcal{L}_\chi$ responsible for off-diagonal correction to $\mu$ matrix :}

Considering the extra scalar $\chi$ with $U(1)_{L_\mu -L_\tau}$ charge 1, the modified Lagrangian with Yukawa-like terms can be written as,
\begin{equation}
\mathcal{L}_{\chi}  \supset   \mu_{e\mu}S^T_{e_L}S_{\mu_L}\chi^{\ast} + \mu_{e\tau}S^T_{e_L}S_{\tau_L}\chi + \mu_{e\mu}S^T_{\mu_L}S_{e_L}\chi^{\ast} + \mu_{e\tau}S^T_{\tau_L}S_{e_L}\chi
\end{equation}
which modifies the structure of the light neutrino mass matrix now looking like, 
\begin{equation}
m_{\nu}^{\text{ELISS}} = \begin{pmatrix}
\frac{a^2 \mu_{ee}}{M_{11}^2} & \frac{ab \mu_{e\mu}}{M_{11}M_{22}} & \frac{ac \mu_{e\tau}}{M_{11}M_{33}} \\
\frac{ab \mu_{e\mu}}{M_{11}M_{22}} & 0 & \frac{bc \mu_{\mu\tau}}{M_{22}M_{33}} \\
\frac{ac \mu_{e\tau}}{M_{11}M_{33}} & \frac{bc \mu_{\mu\tau}}{M_{22}M_{33}} & 0
\end{pmatrix}
\end{equation}

Now, if we consider $M_D$ and $M$ as constant identity mass matrices i.e., $M_D = a \mathbb{I}_{3 \times 3}$ and $M = M_{11} \mathbb{I}_{3 \times 3}$, 
then $m_{\nu}^{\text{ELISS}} \sim \mu$. Since light neutrino mass matrix can be diagonalised by $U_{\text{PMNS}}$ matrix \cite{Tanabashi:2018oca},
\begin{equation}
|U_{\text{PMNS}} | 
\approx
\begin{pmatrix}
0.814 & 0.554 & 0.147 \\
0.329 & 0.572 & 0.717 \\
0.432 & 0.555 & 0.742
\end{pmatrix}
\end{equation}
we can diagonalise $\mu$ by $U_{\text{PMNS}}$ 
and rewrite the mass matrix as (considering the couplings $\mu_{e\mu}=\mu_{e\tau}$ for sake of simplicity), 
\begin{equation}
m_{\nu}^{\prime \text{ELISS}} = \begin{pmatrix}
\frac{a^2 \mu_{ee}}{M_{11}^2} & \frac{a^2 \mu_{e\mu}}{M_{11}^2} & \frac{a^2 \mu_{e\mu}}{M_{11}^2} \\
\frac{a^2 \mu_{e\mu}}{M_{11}^2} &0& \frac{a^2 \mu_{\mu\tau}}{M_{11}^2} \\
\frac{a^2 \mu_{e\mu}}{M_{11}^2} & \frac{a^2 \mu_{\mu\tau}}{M_{11}^2} & 0
\end{pmatrix}
\end{equation}
whose corresponding eigenvalues are $\lbrace \frac{-a^2  \mu_{\mu\tau}}{M_{11}^2}, m_{\nu}^{\prime \text{ELISS} a} \pm m_{\nu}^{\prime \text{ELISS} b} \rbrace$ with 
\begin{center}
$m_{\nu}^{\prime \text{ELISS} a} = \frac{a^2}{2 M_{11}^2} (\mu_{ee}+\mu_{\mu\tau})$, \\
 $m_{\nu}^{\prime \text{ELISS} b} = \frac{a^2}{2 M_{11}^2} \sqrt{\mu_{ee}^2 + 8\mu_{e\mu}^2 -2\mu_{ee}\mu_{\mu\tau}+\mu_{\mu\tau}^2}$
 \end{center}
 \underline{\bf $\mathcal{L}_\chi$ responsible for diagonal correction to $\mu$ matrix :}

Similarly, if we consider $\chi$ with $U(1)_{L_\mu - L_\tau}$ charge 2, then the Lagrangian can be written as,
 \begin{equation}
\mathcal{L}_{\chi}  \supset \mu_{\mu\mu} S^T_{\mu_L}S_{\mu_L}\chi^{\ast} + \mu_{\tau\tau} S^T_{\tau_L}S_{\tau_L}\chi
\end{equation}

Now the modified light neutrino mass matrix in this framework can be 
expressed as,
\begin{equation}
m_{\nu}^{\text{ELISS}} = \begin{pmatrix}
\frac{a^2 \mu_{ee}}{M_{11}^2} & 0 & 0 \\
0 & \frac{b^2 \mu_{\mu\mu}}{M_{22}^2} & \frac{bc \mu_{\mu\tau}}{M_{22}M_{33}} \\
0 & \frac{bc \mu_{\mu\tau}}{M_{22}M_{33}} & \frac{c^2 \mu_{\tau\tau}}{M_{33}^2} 
\end{pmatrix}
\end{equation}
with mass eigenvalues $\lbrace \frac{a^2 \mu_{ee}}{M_{11}^2}, m^{\text{ELISS a}}_{\nu} \pm m^{\text{ELISS b}}_{\nu} \rbrace$ where, 
\begin{center}
$m^{\text{ELISS a}}_{\nu} = \frac{c^2 M_{22}^2 \mu_{\tau\tau} + b^2 M_{33}^2 \mu_{\mu\mu}}{2 M_{22}^2 M_{33}^2} $, \\
 $m^{\text{ELISS b}}_{\nu} = \frac{\sqrt{c^4 M_{22}^4 \mu_{\tau\tau}^2 - 2b^2 c^2 M_{22}^2 M_{33}^2 \mu_{\mu\mu} \mu_{\tau\tau} + b^4 M_{33}^4 \mu_{\mu\mu}^2 + 4b^2 c^2 M_{22}^2 M_{33}^2 \mu_{\mu\tau}^2}}{2 M_{22}^2 M_{33}^2}$
\end{center}

We found that, both the cases i.e with the diagonal as well as off-diagonal corrections to $\mu$-matrix successfully explain current-day neutrino oscillation data at 3$\sigma$ interval of global fit by NuFIT 4.1~\cite{Esteban:2018azc}.

\subsection{Non-standard neutrino interaction via non-unitarity effects in LISS}
\label{subsec:NSI}
\noindent
With the presence of extra sterile neutrinos on top of SM light active neutrinos, the measure of deviation of the neutrino mixing matrix from unitarity is known as non-unitarity effects in leptonic sector which can provide a new window to probe physics beyond Standard Model at present and planned neutrino factories. In the considered framework, left-right inverse seesaw scheme gives non-unitarity effects and thereby can generate non-standard neutrino interaction (NSI). 
The non-unitarity mixing matrix ($\mathbb{N}$) and the measure of deviation from unitarity ($\eta$) can be read as,
\begin{eqnarray}
&&\mathbb{N} \simeq \left(1 - \frac{1}{2} \Theta \Theta^\dagger \right) 
U = \left(1 - \eta \right) U\, , \nonumber \\
&&\eta =  \frac{1}{2} \Theta \Theta^\dagger \, , \mbox{with} \quad \Theta \simeq M_D/M
\end{eqnarray}
where, $U=U_{\rm PMNS}$ being the unitary matrix diagonalizing $m_\nu$ and $\Theta \simeq M_D/M$,
in turn related to the light-heavy neutrino mixing matrix element $V^{\nu \xi}$ to be used next section onwards. This parameter $V^{\nu \xi}$ is a measure of how active light neutrinos mix with heavy sterile neutrinos and can be constrained from non-unitarity effects, NSI effects and muon anomalous $g-2$ etc. The present experimental bounds on the unitarity violation in $e\mu$, $e\tau$, $\mu \tau$, $\tau\tau$ sectors are $|\eta_{e \mu}| < 3.5\times 10^{-5}$, $|\eta_{e\tau}| < 8.0\times 10^{-4}$, $|\eta_{\mu \tau}| < 5.1\times 10^{-3}$  and $|\eta_{\tau\tau}| < 2.7\times 10^{-3}$ \cite{Antusch:2008tz,Awasthi:2013ff} respectively. For a more detailed study on low energy LFV processes $\mu\to e\gamma$, $\mu\to eee$ and $\mu\to e$~conversion in nuclei due to non-unitarity effects readers can refer \cite{Deppisch:2012vj,Blennow:2016jkn}. Through charged current interaction and expressing light active neutrinos in terms of mass eigenstates including light as well as sterile neutrinos, the heavy sterile neutrinos ($\xi$) couple to gauge sector of SM which eventually create non-standard interaction (NSI) for neutrinos.  In NSI effects for a given non-unitarity lepton mixing matrix $\mathbb{N}$, the vacuum neutrino oscillation probability ${P}_{\alpha \beta}$ can be expressed as \cite{Ohlsson:2008gx},
\begin{eqnarray}
P_{\alpha\beta} &=&  \sum_{i,j} {\cal F}^i_{\alpha\beta} {\cal
F}^{j*}_{\alpha\beta} - 4 \sum_{i>j} {\rm Re} ({\cal
F}^i_{\alpha\beta} {\cal F}^{j*}_{\alpha\beta} )\sin^2\!\left(
\frac{\Delta m^{2}_{ij}L}{4E}\right) \nonumber \\ &+& 2
\sum_{i>j}{\rm Im} ( {\cal F}^i_{\alpha\beta} {\cal
F}^{j*}_{\alpha\beta} ) \sin\left(\frac{ \Delta m^{2}_{ij} L}{2
E}\right) \ ,
\label{eq:P}
\end{eqnarray}
where $\Delta m^{2}_{ij} \equiv m^2_i - m^2_j$ are the neutrino
mass-squared differences and ${\cal F}^i$ are defined by
\begin{eqnarray}\label{eq:F}
{\cal F}^i_{\alpha\beta} \equiv \sum_{\gamma ,\rho} ( R^*)_{\alpha
\gamma } ( R^*)^{-1}_{\rho \beta } U^*_{\gamma i} U_{\rho i} \,.
\end{eqnarray}
Here, the normalized non-unitary factor in terms of $\eta$ parameters is given by
\begin{eqnarray}\label{eq:R}
R_{\alpha\beta} \equiv \frac{(1-\eta)_{\alpha\beta}}
{\left[(1-\eta)(1-\eta^\dagger)\right]_{\alpha\alpha}} \ .
\end{eqnarray}
The mass parameters $M_D$ and $M$ are found to be diagonal in the present framework with $U(1)_{L_\mu-L_\tau}$ symmetry. This makes few elements of $\eta$ negligible and hence restricts $R_{\alpha\beta}$. 
The neutrino factory will provide excellent sensitivity to probe these non-standard interaction effects and for more details about NSI in inverse seesaw mechanism, one may read refs\cite{Ohlsson:2008gx,FernandezMartinez:2007ms,Malinsky:2009gw,Meloni:2009cg,Malinsky:2009df,Babu:2019mfe,Blennow:2016jkn}. 
Presence of heavy neutral fermions leads to non-unitarity effects and it has been shown that one may constrain these heavy fermions by studying their impact by adding few effective
operators $\mathcal{O}^{d}$ of dimension $d>4$ to the interaction Lagrangian\cite{Meloni:2009cg}.
We also skip the detailed phenomenology of LISS (interested readers may refer ~\cite{Dev:2009aw,Awasthi:2013ff,Pritimita:2016fgr,Majee:2008mn,LalAwasthi:2011aa,Meloni:2009cg,Malinsky:2009gw,Ohlsson:2008gx,Malinsky:2009df}) in the context of  cLFV, non-unitarity effects, $0\nu\beta\beta$, LNV at collider. Rather, we intend to explore the implications of 
light-heavy neutrino mixing $V^{\nu\xi}$ with purely left-handed currents to new physics contributions to muon $g-2$ anomaly in the following section.

\section{Prediction on Muon $(g-2)$ anomaly}
\label{sec:muon_anomaly_prediction}
For a comprehensive review on new physics scenarios explaining muon $(g-2)$ anomaly one may refer \cite{Jegerlehner:2009ry,Lindner:2016bgg, Queiroz:2014zfa}. Most of these works 
predict that new light gauge bosons and light neutral scalars are good candidates for addressing the anomaly since they contribute positively to $\Delta a_\mu$.
In our model, new contributions to muon $(g-2)$ anomaly arise from the interactions of;
\begin{itemize}
 \item singly charged gauge bosons with heavy neutral fermions,
 \item neutral vector boson with singly charged fermions,
 \item singly charged scalars with neutral fermion,
 \item neutral scalars with muons,
 \item extra light new gauge boson $Z_{\mu\tau}$ with muons.
\end{itemize}

In the following we study analytically all these new physics contributions to $\Delta a_\mu$ and numerically estimate the 
individual contributions in the next section.
Notably, for the calculation of $\Delta a_\mu$ we neglect the flavor mixing as they give negligible correction to the anomaly \cite{Queiroz:2014zfa}. 
Another important point to recall here is that inverse seesaw mechanism which explains neutrino mass in this model also allows large light-heavy 
neutrino mixing due to which the contribution coming from the charged gauge boson interaction with heavy neutral fermion becomes sizeable.

\subsection{Gauge boson contribution}
Before moving on to the Feynman diagrams mediated by gauge bosons, we write the basic charged current(CC) interaction Lagrangian for leptons  within left-right theories.
\begin{equation}
\mathcal{L}^l_\text{cc} = \sum_{\alpha = e, \mu , \tau} \left[  \frac{g_L}{\sqrt{2}} \overline{\ell}_{\alpha L} 
\gamma_\beta \ell_{\alpha L}W_L^\beta + \frac{g_R}{\sqrt{2}} \overline{\ell}_{\alpha R} \gamma_\beta \ell_{\alpha R}W_R^\beta \right] +\text{h.c.}
\label{eq:cclrsm}
\end{equation}
For Inverse Seesaw (ISS) mechanism \cite{Awasthi:2013ff,Dev:2009aw}, the flavour eigenstates 
$\nu_L$ and $N_R$ can be expressed in terms of admixture of mass eigenstates ($\nu_i$, $\xi_j$) as follows,
\begin{equation}
\nu_{\mu L} = V_{\mu i}^{\nu \nu} \nu_i + V_{\mu j}^{\nu \xi} \xi_j
\label{lh_mbasis_iss}
\end{equation}
\begin{equation}
N_{\mu R} = V_{\mu i}^{N \nu} \nu_i + V_{\mu j}^{N \xi} \xi_j
\label{rh_mbasis_iss}
\end{equation}
where $i =1,2,3$ goes over physical states for light neutrinos and $j=1, 2, ....., 6$ runs over heavy states forming three pairs of pseudo-Dirac neutrinos. Using eq.(\ref{lh_mbasis_iss}) in the charged current interaction lagrangian given in  eq.\ref{eq:cclrsm}, we present the vector and axial vector couplings ($g_{v}$ and $g_a$) in Table.\ref{xi_coupling_ISS}.\\
\begin{table}[h!]
\centering
\begin{tabular}{|c|c|c||c|c|c|}
\hline
  Interaction Vertex & $g_{v2} = - g_{a2}$ & Interaction Vertex & $g_{v1} = g_{a1}$
           \\[3mm]
\hline \hline 
$\overline{\nu}_1 \mu W_L^+$   & $\frac{g_L}{2 \sqrt{2}}V_{\mu 1}^{\nu\nu \ast}$  
& $\overline{\nu}_1 \mu W_R^+$   & $\frac{g_R}{2 \sqrt{2}}V_{\mu 1}^{N\nu \ast}$ 
          \\[2mm]
 \hline
 $\overline{\nu}_2 \mu W_L^+$   & $\frac{g_L}{2 \sqrt{2}}V_{\mu 2}^{\nu\nu \ast}$ 
 & $\overline{\nu}_2 \mu W_R^+$   & $\frac{g_R}{2 \sqrt{2}}V_{\mu 1}^{N\nu \ast}$
          \\[2mm]
 \hline
 $\overline{\nu}_3 \mu W_L^+$   & $\frac{g_L}{2 \sqrt{2}}V_{\mu 3}^{\nu\nu \ast}$
 & $\overline{\nu}_3 \mu W_R^+$   & $\frac{g_R}{2 \sqrt{2}}V_{\mu 1}^{N\nu \ast}$
          \\[2mm]
 \hline
 $\overline{\xi}_1 \mu W_L^+$   & $\frac{g_L}{2 \sqrt{2}}V_{\mu 1}^{\nu \xi \ast}$
 & $\overline{\xi}_1 \mu W_R^+$   & $\frac{g_R}{2 \sqrt{2}}V_{\mu 1}^{N \xi \ast}$
           \\[2mm]
 \hline
  $\overline{\xi}_2 \mu W_L^+$   & $\frac{g_L}{2 \sqrt{2}}V_{\mu 2}^{\nu \xi \ast}$
  & $\overline{\xi}_2 \mu W_R^+$   & $\frac{g_R}{2 \sqrt{2}}V_{\mu 2}^{N \xi \ast}$
          \\[2mm]
 \hline
  $\overline{\xi}_3 \mu W_L^+$   & $\frac{g_L}{2 \sqrt{2}}V_{\mu 3}^{\nu \xi \ast}$
  & $\overline{\xi}_3 \mu W_R^+$   & $\frac{g_R}{2 \sqrt{2}}V_{\mu 3}^{N \xi \ast}$
          \\[2mm]
 \hline
  $\overline{\xi}_4 \mu W_L^+$   & $\frac{g_L}{2 \sqrt{2}}V_{\mu 4}^{\nu \xi \ast}$
  & $\overline{\xi}_4 \mu W_R^+$   & $\frac{g_R}{2 \sqrt{2}}V_{\mu 4}^{N \xi \ast}$
          \\[2mm]
 \hline
 $\overline{\xi}_5 \mu W_L^+$   & $\frac{g_L}{2 \sqrt{2}}V_{\mu 5}^{\nu \xi \ast}$
 & $\overline{\xi}_5 \mu W_R^+$   & $\frac{g_R}{2 \sqrt{2}}V_{\mu 5}^{N \xi \ast}$
          \\[2mm]
 \hline
 $\overline{\xi}_6 \mu W_L^+$   & $\frac{g_L}{2 \sqrt{2}}V_{\mu 6}^{\nu \xi \ast}$
 & $\overline{\xi}_6 \mu W_R^+$   & $\frac{g_R}{2 \sqrt{2}}V_{\mu 6}^{N \xi \ast}$
          \\[2mm]
 \hline
\end{tabular}
\caption{Relevant vector and axial vector couplings for muon with $W_L,W_R$ gauge bosons and physical neutral fermion states within  the inverse seesaw (ISS) scenario.}
\label{xi_coupling_ISS}
\end{table}

In inverse seesaw scheme, the light neutrinos are Majorana in nature while heavy neutrinos are pseudo-Dirac. Alternatively, in extended inverse seesaw scenario (EISS) \cite{Pritimita:2016fgr, Awasthi:2013ff} both light neutrino $\nu_L$ as well as heavy neutrinos $S_L, N_R$ are purely Majorana in nature. Thus, the flavour eigenstates $\nu_L$ and $N_R$ 
can be expressed in terms of admixture of mass eigenstates ($\nu_i$, $S_i$, $N_i$) in the following way,
\begin{equation}
\nu_{\mu L} = V_{\mu i}^{\nu \nu} \nu_i + V_{\mu i}^{\nu S} S_i + V_{\mu i}^{\nu N} N_i
\label{lh_mbasis_eiss}
\end{equation}
\begin{equation}
N_{\mu R} = V_{\mu i}^{N \nu} \nu_i + V_{\mu i}^{N S} S_i + V_{\mu i}^{N N} N_i
\label{rh_mbasis_eiss}
\end{equation}
where $i =1,2,3$ goes over physical states. For EISS we present the vector and axial vector couplings in Table.\ref{xi_coupling_EISS}.\\
\begin{table}[h!]
\centering
\begin{tabular}{|c|c|c||c|c|c|}
\hline
  Interaction Vertex & $g_{v2} = - g_{a2}$ & Interaction Vertex & $g_{v1} = g_{a1}$
           \\[3mm]
\hline \hline 
$\overline{\nu}_1 \mu W_L^+$   & $\frac{g_L}{2 \sqrt{2}}V_{\mu 1}^{\nu\nu \ast}$ 
& $\overline{\nu}_1 \mu W_R^+$   & $\frac{g_R}{2 \sqrt{2}}V_{\mu 1}^{N\nu \ast}$
          \\[2mm]
 \hline
 $\overline{\nu}_2 \mu W_L^+$   & $\frac{g_L}{2 \sqrt{2}}V_{\mu 2}^{\nu\nu \ast}$
 & $\overline{\nu}_2 \mu W_R^+$   & $\frac{g_R}{2 \sqrt{2}}V_{\mu 1}^{N\nu \ast}$
          \\[2mm]
 \hline
 $\overline{\nu}_3 \mu W_L^+$   & $\frac{g_L}{2 \sqrt{2}}V_{\mu 3}^{\nu\nu \ast}$
 & $\overline{\nu}_3 \mu W_R^+$   & $\frac{g_R}{2 \sqrt{2}}V_{\mu 1}^{N\nu \ast}$
          \\[2mm]
 \hline
 $\overline{S}_1 \mu W_L^+$   & $\frac{g_L}{2 \sqrt{2}}V_{\mu 1}^{\nu S \ast}$
 & $\overline{S}_1 \mu W_R^+$   & $\frac{g_R}{2 \sqrt{2}}V_{\mu 1}^{N S \ast}$
          \\[2mm]
 \hline
  $\overline{S}_2 \mu W_L^+$   & $\frac{g_L}{2 \sqrt{2}}V_{\mu 2}^{\nu S \ast}$
  & $\overline{S}_2 \mu W_R^+$   & $\frac{g_R}{2 \sqrt{2}}V_{\mu 2}^{N S \ast}$
          \\[2mm]
 \hline
  $\overline{S}_3 \mu W_L^+$   & $\frac{g_L}{2 \sqrt{2}}V_{\mu 3}^{\nu S \ast}$
  & $\overline{S}_3 \mu W_R^+$   & $\frac{g_R}{2 \sqrt{2}}V_{\mu 3}^{N S \ast}$
          \\[2mm]
 \hline
  $\overline{N}_1 \mu W_L^+$   & $\frac{g_L}{2 \sqrt{2}}V_{\mu 1}^{\nu N \ast}$
  & $\overline{N}_1 \mu W_R^+$   & $\frac{g_R}{2 \sqrt{2}}V_{\mu 1}^{N N \ast}$
          \\[2mm]
 \hline
 $\overline{N}_2 \mu W_L^+$   & $\frac{g_L}{2 \sqrt{2}}V_{\mu 2}^{\nu N \ast}$ 
 & $\overline{N}_2 \mu W_R^+$   & $\frac{g_R}{2 \sqrt{2}}V_{\mu 2}^{N N \ast}$
          \\[2mm]
 \hline
 $\overline{N}_3 \mu W_L^+$   & $\frac{g_L}{2 \sqrt{2}}V_{\mu 3}^{\nu N \ast}$
 & $\overline{N}_3 \mu W_R^+$   & $\frac{g_R}{2 \sqrt{2}}V_{\mu 3}^{N N \ast}$
          \\[2mm]
 \hline
\end{tabular}
\caption{Relevant vector and axial vector couplings for muon with $W_L,W_R$ gauge bosons and physical neutral fermion states within  the extended inverse seesaw (EISS) scenario.}
\label{xi_coupling_EISS}
\end{table}

\begin{figure}[h]
\centering
\includegraphics[scale=0.8]{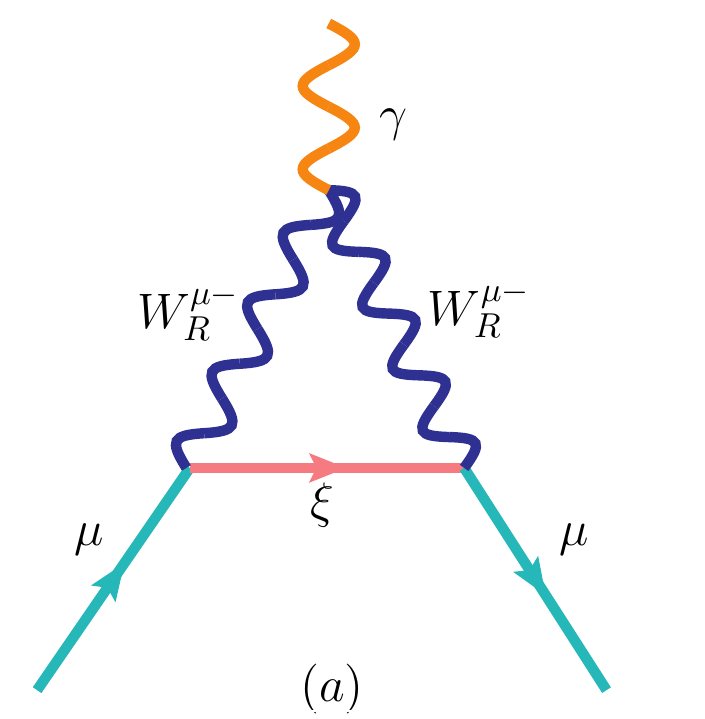}\qquad
\includegraphics[scale=0.8]{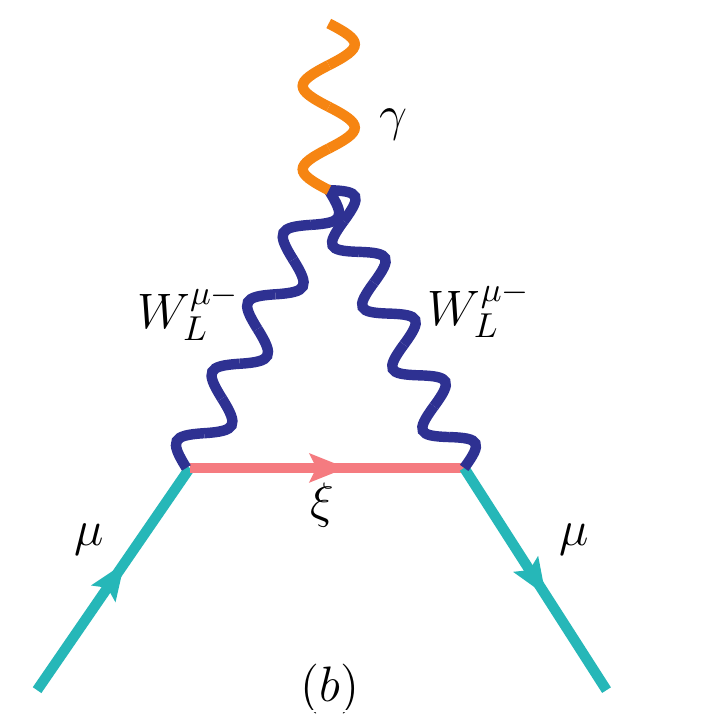}
\caption{Feynman diagrams for the interaction of singly charged vector bosons: in left-panel due to the mediation of singly charged right-handed gauge boson $W_R$ with heavy neutrinos and in right-panel due to the mediation of singly charged left-handed gauge boson $W_L$ with exchange of heavy neutrinos. The $W_L$ mediated diagram with exchange of heavy neutrinos gives sizeable contribution in ISS scheme.}
\label{W-mediation}
\end{figure}

The diagrams in Fig.\ref{W-mediation} are mediated by singly charged right-handed and left-handed gauge bosons $W_R$, $W_L$ interacting with muons. 
Here $\xi$ represents the heavy neutrino states in mass basis within the inverse seesaw framework. 
$W_L$ can interact with heavy right-handed neutrino due to inverse seesaw mechanism in the model, and we find out in the next section 
that the most significant contribution to muon anomaly comes from this channel.
For a detailed discussion on the contributions arising from singly charged vector bosons one may refer \cite{Cogollo:2012ek, Cao:2012ng, Dong:2013ioa, Dong:2013wca}.

\noindent
\underline{\bf Fig.1(a): Contribution due to $W_R$ mediation; $\Delta a_\mu (\xi, W_{R})$}\\
For calculating its contribution, we start by sorting out the relevant interaction terms for this diagram.
\begin{equation}
\mathcal{L_\text{int}} = g_{v1} W^+_{R \mu} \overline{\nu_\mu} \gamma^\mu \mu + g_{a1} W^+_{R \mu} \overline{\nu_\mu} \gamma^\mu \gamma^5 \mu + h.c.
\end{equation}
The contribution arising from this diagram to the anomalous magnetic moment can be determined by the following expression.
\begin{equation}
\Delta a_\mu (\xi, W_{R}) \simeq \frac{1}{8\pi^2}\frac{m^2_\mu}{m^2_{W_{R}}} \int_0^1 dx \dfrac{g_{v1}^2 P_{v1}(x) + g_{a1}^2 P_{a1}(x)}{\epsilon^2 \lambda^2 (1-x) (1- \epsilon^{-2} x) +x}
\end{equation}
where, $m_\mu$ is the mass of muon, $m_{W_R}$ is the mass of right-handed charged gauge boson $W_R$, $ \epsilon \equiv \left(\frac{m_{\nu_\mu}}{m_{\mu}}\right),~ \lambda \equiv \left(\frac{m_\mu}{m_{W_R}}\right)$ , and
\begin{center}
$ P_{v1}(x) = 2x^2(1+x -2 \epsilon) - \lambda^2 (1- \epsilon)^2 x(1-x)(x+ \epsilon) $\\
$ P_{a1}(x) = 2x^2(1+x +2 \epsilon) - \lambda^2 (1 + \epsilon)^2 x(1-x)(x - \epsilon) $
\end{center}
After simplifying the integration the expression can be rewritten as (we will neglect the terms containing $\epsilon$ and $\lambda$ in the expression of muon anomaly $\Delta a_\mu$ onwards (except neutral scalar sector to be discussed in next subsections) as they are really tiny corrections),
\begin{equation}
\Delta a_\mu (\xi, W_R) \simeq \frac{1}{4\pi^2}\frac{m^2_\mu}{m^2_{W_R}}\left[|g_{v1}^\mu|^2 \left(\frac{5}{6}  \right) + |g_{a1}^\mu|^2 \left(\frac{5}{6}  \right) \right];~~~\text{with}~~~ m_{W_R} \gg m_\mu.
\label{npc_wr}
\end{equation}
Here we have, $|g_{v1}| = |g_{a1}| = \frac{g_R}{2\sqrt{2}}$ (as given in Table\ref{xi_coupling_ISS} with $\mathcal{O}$(1) neutrino mixing) and with these values we can rewrite Eq.\ref{npc_wr} as,
\begin{equation}
\Delta a_\mu (W_R) \simeq 2.3 \times 10^{-11} \left( \frac{g_R}{g_L}\right)^2 \left(  \frac{1~\text{TeV}}{m_{W_R}}\right)^2 \sum_{i=1,..,6} |V_{\mu i}^{N \xi}|^2
\end{equation}
\noindent
\underline{\bf Fig.1(b): Contribution due to $W_L$ mediation with light-heavy neutrino mixing; $\Delta a_\mu (\xi, W_L)$}\\ Similar as 1(a) the relevant interaction terms for this diagram.
\begin{equation}
\mathcal{L_\text{int}} = g_{v1} W^+_{L \mu} \overline{\nu_\mu} \gamma^\mu \mu + g_{a1} W^+_{L \mu} \overline{\nu_\mu} \gamma^\mu \gamma^5 \mu + h.c.
\end{equation}
So, for $W_L$ interacting with heavy neutrino the contribution to muon anomalous magnetic moment can be expressed as,
\begin{equation}
\Delta a_\mu (\xi, W_L) \simeq \frac{1}{4\pi^2}\frac{m^2_\mu}{m^2_{W_L}}\left[|g_{v2}^\mu|^2 \left(\frac{5}{6}  \right) + |g_{a2}^\mu|^2 \left(\frac{5}{6}  \right) \right];~~~\text{with}~~~ m_{W_L} \gg m_\mu.
\label{npc_wl}
\end{equation} 
Using the couplings for this interaction given in Table \ref{xi_coupling_ISS} we can rewrite Eq.\ref{npc_wl} as,
\begin{equation}
\Delta a_\mu (\xi, W_L) \simeq 9.06 \times 10^{-9}~g_L^2  \sum_{i=1,..,6} |V_{\mu i}^{\nu \xi}|^2
\end{equation}
Since the ISS scenario allows large mixing between light and heavy neutrinos, moving from flavor to mass basis we can see that for $\mathcal{O}(0.1)$ light-heavy neutrino mixing, heavy neutrinos with mass $\sim $ few GeV play a significant role in context of muon $g-2$ anomaly by interacting with $W_L$. Also, in the next section we will see that this gives positive and significant contribution to $\Delta a_\mu$.

\noindent
\underline{\bf Fig.2: Contribution due to $Z_R$ mediation; $\Delta a_\mu (Z_{R})$}\\ 
\begin{figure}[h]
\centering
\includegraphics[scale=0.8]{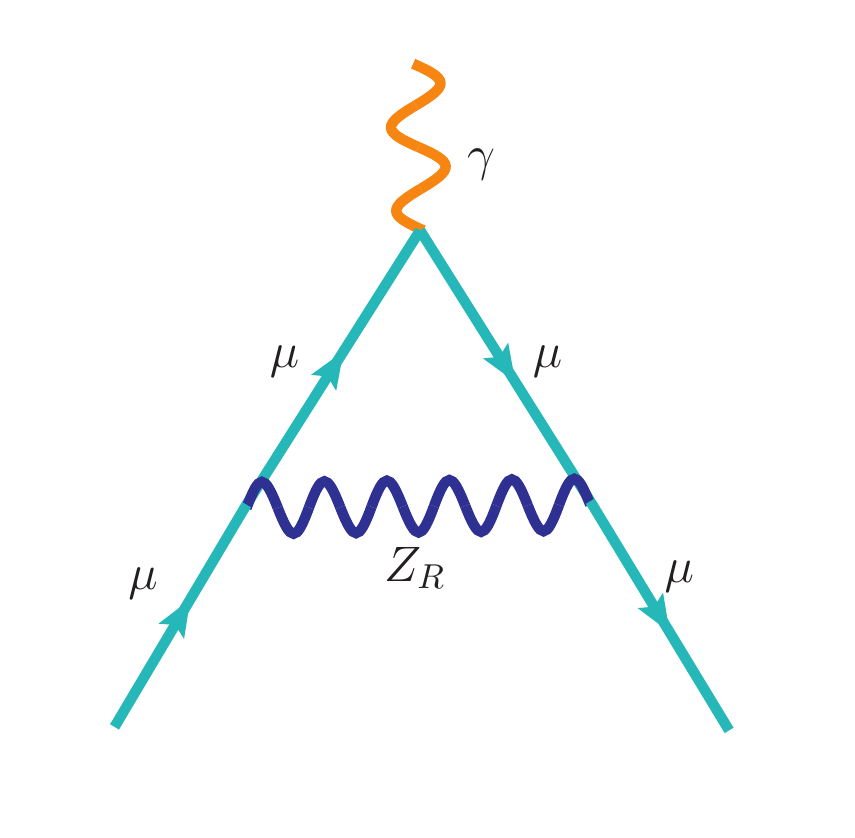}
\caption{Feynman diagram for muon anomalous $g-2$ contribution arising from the mediation of right-handed neutral gauge boson  $Z_R$ with muons.}
\label{ZR-mediation}
\end{figure}
The new contribution for muon anomalous $g-2$ arising from exchange of right-handed neutral gauge boson $Z_R$, as shown in Fig.\ref{ZR-mediation}, is derived from the neutral current interaction as,
\begin{equation}
\bar{\mu}\gamma_\beta \partial^\beta \mu +i\frac{g_L}{\sqrt{1- \delta \text{tan}^2 \theta_W}} \overline{\mu} \gamma_\beta (g_{v} - g_{a} \gamma^5) \mu Z_R^\beta
\label{nvector}
\end{equation}
with the couplings
\begin{center}
$g_{v} =  \frac{1}{4}\left[ 3 \delta  \text{tan}^2 \theta_W - 1 \right]$ \\
$g_{a} =   \frac{1}{4}\left[ 1-  \delta  \text{tan}^2 \theta_W\right] $
\end{center}
where $\delta = \frac{g^2_L}{g^2_R}$ and $\theta_W$ is the Weinberg angle. The Lagrangian for the charged fermions which interact with the SM leptons via a neutral vector boson ($ Z_{R} $) can be written as
\begin{equation}
\mathcal{L_\text{int}} = g_{v3} Z_{R\mu} \overline{\mu} \gamma^\mu \mu + g_{a3} Z_{R\mu} \overline{\mu} \gamma^\mu \gamma^5 \mu + h.c.
\label{int_nv}
\end{equation}
Using Eq.\ref{int_nv} the contribution arising from $Z_R$ to the muon anomalous magnetic moment can be expressed as,
\begin{equation}
\Delta a_\mu (Z_{R}) \simeq \frac{1}{8\pi^2}\frac{m^2_\mu}{m^2_{Z_R}} \int_0^1 dx \dfrac{g_{v3}^2 P_{v3}(x) + g_{a3}^2 P_{a3}(x)}{(1-x) (1- \lambda^2  x) +\lambda^2 x}
\end{equation}
with $\lambda \equiv \left(\frac{m_\mu}{m_{Z_R}}\right)$, and
\begin{center}
$ P_{v3}(x) = 2x^2(1-x) $\\
$ P_{a3}(x) = 2x(1-x)(x-4)- 4 \lambda^2 x^3 $
\end{center}
By simplifying the integrations the contribution is found to be,
\begin{equation}
\Delta a_\mu (Z_R) \simeq -\frac{1}{4\pi^2}\frac{m^2_\mu}{m^2_{Z_R}}\left[ \left(-\frac{1}{3}\right)|g_{v3}^\mu|^2  + \left(\frac{5}{3}\right)|g_{a3}^\mu|^2 \right];~~~\text{with}~~~ m_{Z_R} \gg m_\mu.
\end{equation}
where the couplings $g_{v3}$, $g_{a3}$ are same as $g_v$, $g_a$ respectively as in Eq.\ref{nvector} and depending on the values of these vector and axial couplings 
the contribution can be either positive or negative.
\subsection{Scalar sector contribution}
\begin{figure}[h]
\centering
\includegraphics[scale=0.8]{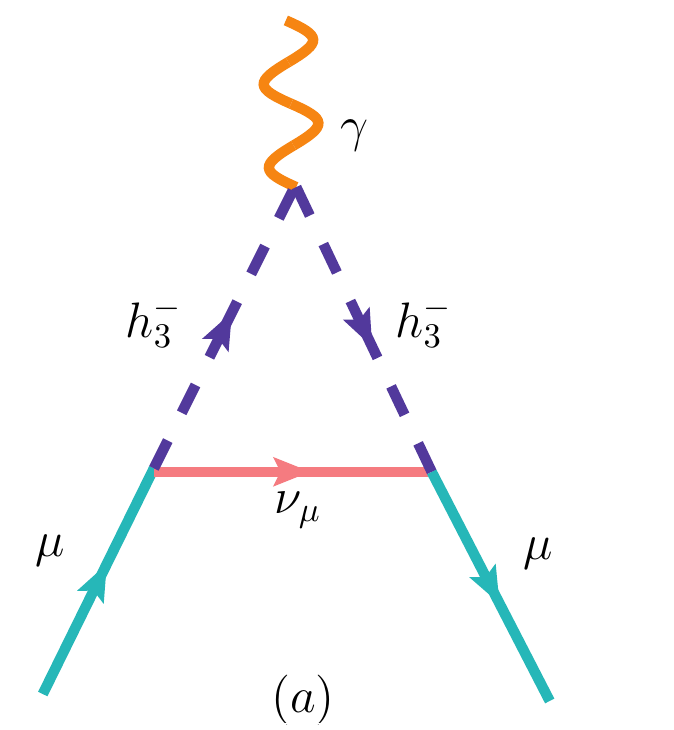}\qquad
\includegraphics[scale=0.8]{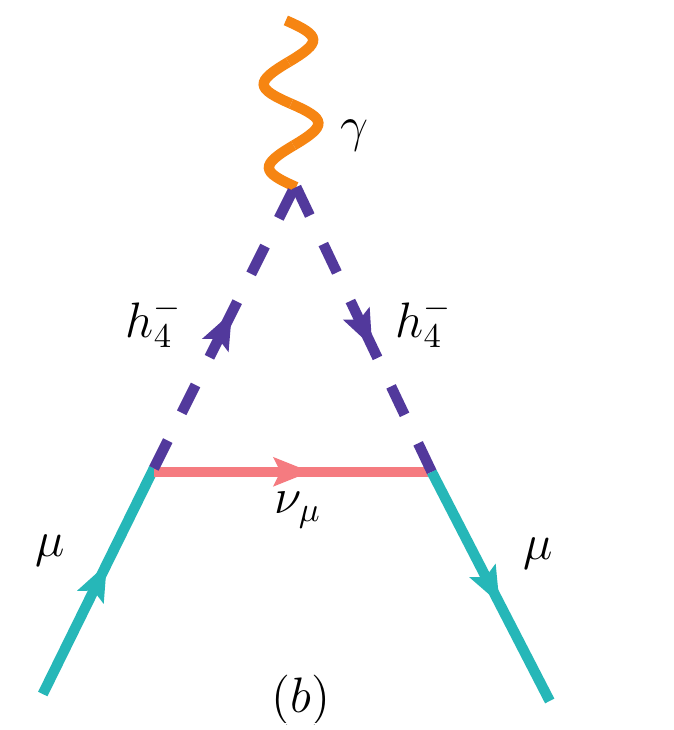}
\caption{Feynman diagrams for the interaction of singly charged scalars $h_3^-, h_4^-$ with muons contributing to the muon anomalous $g-2$.}
\label{charged_scalar}
\end{figure}

\begin{figure}[h]
\centering
\includegraphics[scale=0.8]{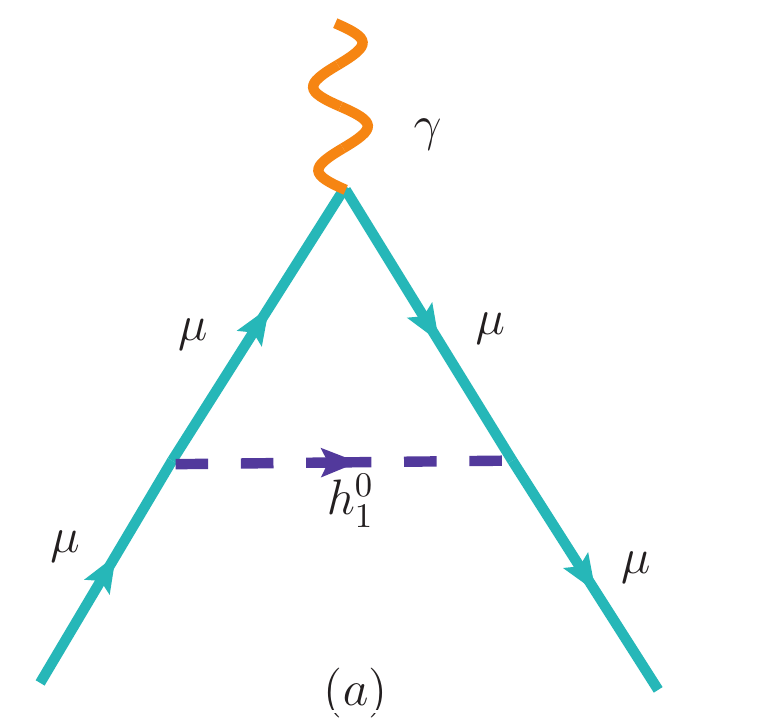}\qquad
\includegraphics[scale=0.8]{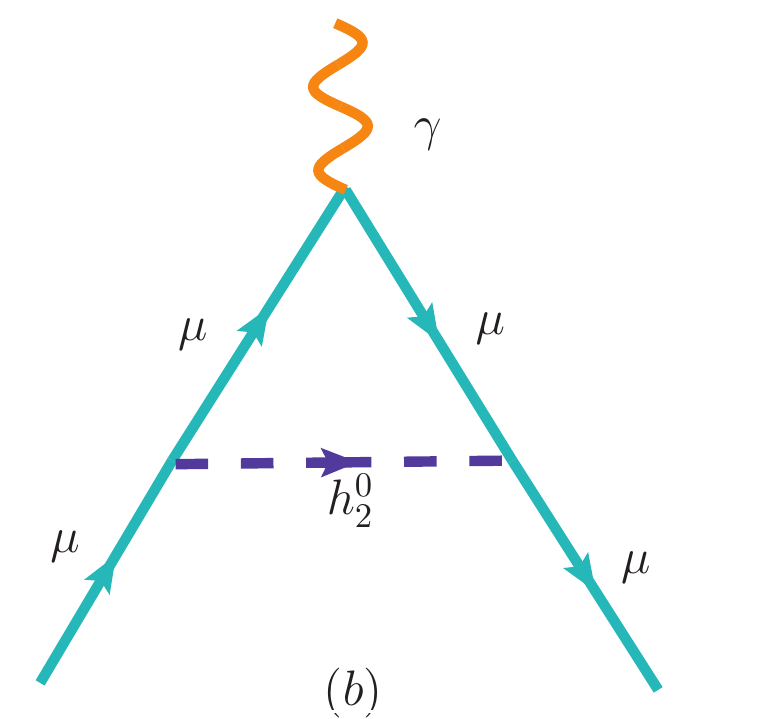}
\caption{Feynman diagrams for the interaction of neutral CP-even scalars $h_1^0, h_2^0$ with muons.}
\label{neutral_CP_even_scalar}
\end{figure}

\begin{figure}[h]
\centering
\includegraphics[scale=0.8]{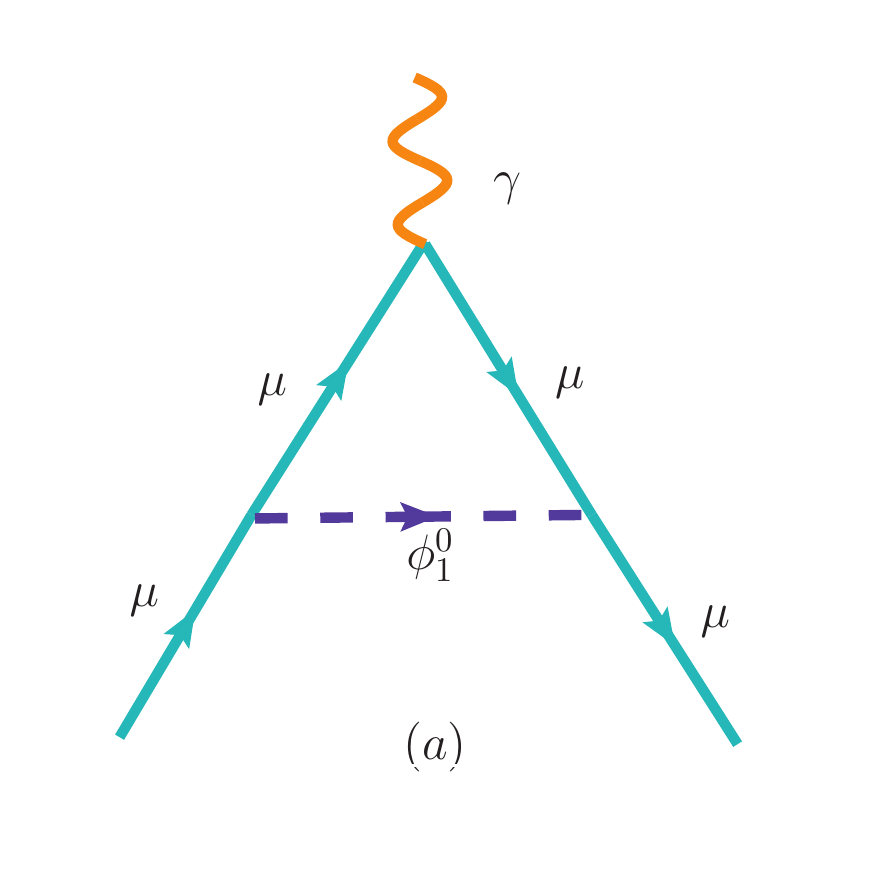}\qquad
\includegraphics[scale=0.8]{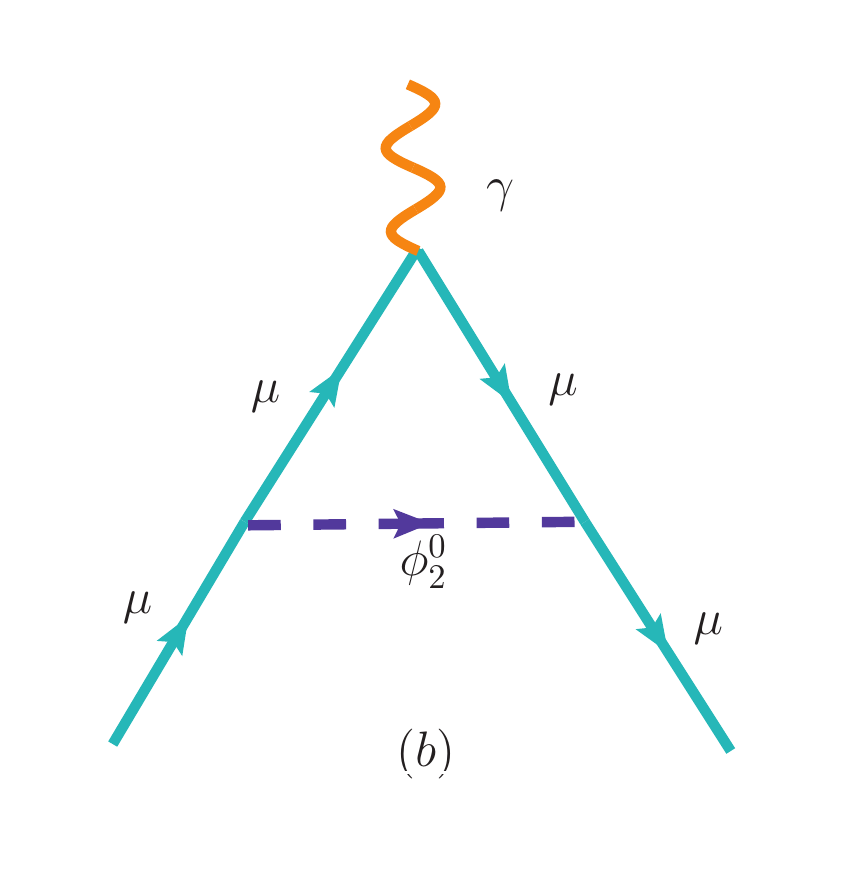}
\caption{Feynman diagrams for the interaction of neutral CP-odd scalars $\phi_1^0, \phi_2^0$ with muons.}
\label{neutral_CP_odd_scalar}
\end{figure}

The Yukawa Lagrangian involving scalars can be written as,
\begin{equation}
\mathcal{L}_\text{Yuk} =  \overline{\ell}_L(Y_{22}\Phi + \tilde{Y}_{22}\tilde{\Phi}) \ell_R  +\overline{\ell}_R(Y_{22}\Phi^{\ast} + \tilde{Y}_{22}\tilde{\Phi}^{\ast}) \ell_L
\end{equation}
where the scalar bidoublet $\Phi$ contains two charged scalars $h_3^-,h_4^-$, two neutral CP-even scalars $h_1^0,h_2^0$ and two neutral CP-odd scalars $\phi_1^0,\phi_2^0$ as 
\begin{center}
$ \Phi =
\begin{pmatrix}
v_1 +h_1^0 + i\phi_1^0 & h_3^+ \\
h_4^- & v_2 +h_2^0 + i\phi_2^0
\end{pmatrix}$ 
\end{center}
and 
\begin{center}
$\tilde{\Phi} = \sigma^2 \Phi^{\ast} \sigma^2 = 
\begin{pmatrix}
v_2 +h_2^0 - i\phi_2^0 & -h_4^+ \\
-h_3^- & v_1 +h_1^0 - i\phi_1^0
\end{pmatrix}$
\end{center}
The Feynman diagrams of these scalars interacting with muons are shown in Figures \ref{charged_scalar}, \ref{neutral_CP_even_scalar}, \ref{neutral_CP_odd_scalar} respectively. 
We later find out in Sec \ref{sec:results} that among these only the 
neutral CP-even scalars $h_1^0,h_2^0$ contribute positively to $\Delta a_\mu$.
Now by considering only muon family with  
\begin{center}
$\ell_L = 
\begin{pmatrix}
\nu_{\mu L} \\
\mu_L
\end{pmatrix}$, $\ell_R = 
\begin{pmatrix}
N_{\mu R} \\
\mu_R
\end{pmatrix}$, 
\end{center}
the expanded Yukawa Lagrangian can be written as,
\begin{align}
\mathcal{L}_\text{Yuk} & =  \left[\overline{\nu}_\mu \left[ Y_{22}(v_1 +h_1^0 + i\phi_1^0) +   \tilde{Y}_{22} (v_2 +h_2^0 - i\phi_2^0) \right] N_\mu +  \overline{\nu}_\mu \left[ Y_{22} h_3^+ -  \tilde{Y}_{22}h_4^+ \right] \mu \right] \frac{(1+\gamma_5)}{2} \nonumber \\
& + \left[\overline{\mu} \left[ Y_{22} h_4^- -  \tilde{Y}_{22}h_3^- \right] N_\mu + \overline{\mu} \left[ Y_{22}(v_2 +h_2^0 + i\phi_2^0) +   \tilde{Y}_{22} (v_1 +h_1^0 - i\phi_1^0) \right] \mu \right] \frac{(1+\gamma_5)}{2} \nonumber \\
& + \left[ \overline{N}_\mu \left[ Y_{22}(v_1 +h_1^0 - i\phi_1^0) +   \tilde{Y}_{22} (v_2 +h_2^0 + i\phi_2^0) \right] \nu_\mu +  \overline{N}_\mu \left[ Y_{22} h_3^- -  \tilde{Y}_{22}h_4^- \right] \mu \right] \frac{(1-\gamma_5)}{2}\nonumber \\
& + \left[\overline{\mu} \left[ Y_{22} h_4^+ -  \tilde{Y}_{22}h_3^+ \right] \nu_\mu + \overline{\mu} \left[ Y_{22}(v_2 +h_2^0 - i\phi_2^0) +   \tilde{Y}_{22} (v_1 +h_1^0 + i\phi_1^0) \right] \mu \right] \frac{(1-\gamma_5)}{2}
\label{scalar_mbasis}
\end{align}
The relevant terms in the Yukawa Lagrangian for the Feynman diagrams given in Fig.3 are as follows,
\begin{equation}
\mathcal{L}_\text{Yuk}(h_3^+, h_4^+) = \overline{\nu}_\mu \left[ Y_{22} h_3^+ -  \tilde{Y}_{22}h_4^+ \right] \mu \frac{(1+\gamma_5)}{2} + \overline{\mu} \left[ Y_{22} h_4^+ -  \tilde{Y}_{22}h_3^+ \right] \nu_\mu  \frac{(1-\gamma_5)}{2}
\label{charged_scalar_fbasis}
\end{equation}
The same equation can be written in mass basis using \ref{lh_mbasis_iss} as,
\begin{align}
\mathcal{L}^{\text{mass}}_\text{Yuk}(h_3^+, h_4^+) & = [V_{\mu 1}^{\nu\nu \ast}\overline{\nu}_1+V_{\mu 2}^{\nu\nu \ast}\overline{\nu}_2+V_{\mu 3}^{\nu\nu \ast}\overline{\nu}_3 + V_{\mu 1}^{\nu S \ast}\overline{S}_1+V_{\mu 2}^{\nu S \ast}\overline{S}_2+V_{\mu 3}^{\nu S \ast}\overline{S}_3+V_{\mu 1}^{\nu N \ast}\overline{N}_1 \nonumber \\
& +V_{\mu 2}^{\nu N \ast}\overline{N}_2+V_{\mu 3}^{\nu N \ast}\overline{N}_3 ] \left[ Y_{22} h_3^+ -  \tilde{Y}_{22}h_4^+ \right] \mu \frac{(1+\gamma_5)}{2} + \overline{\mu} \left[ Y_{22} h_4^+ -  \tilde{Y}_{22}h_3^+ \right] \nonumber \\
& [V_{\mu 1}^{\nu\nu }\nu_1+V_{\mu 2}^{\nu\nu}\nu_2+V_{\mu 3}^{\nu\nu} \nu_3 + V_{\mu 1}^{\nu S }S_1+V_{\mu 2}^{\nu S}S_2+V_{\mu 3}^{\nu S }S_3+V_{\mu 1}^{\nu N }N_1+V_{\mu 2}^{\nu N }N_2 \nonumber \\
& +V_{\mu 3}^{\nu N}N_3]  \frac{(1-\gamma_5)}{2}
\label{charged_scalar_mbasis}
\end{align}
The diagrams in Fig.\ref{charged_scalar} represent the interactions mediated by singly charged scalars $h_3^-$ and $h_4^-$. \\
\underline{\bf Fig.3(a): Contribution due to charged scalar, $h^+_3$ mediation; $\Delta a_\mu (h^+_{3})$} \\
The relevant interaction terms involving singly charged scalar with scalar coupling ($g_{s1}$) and pseudo-scalar coupling ($g_{p1}$) are given by,
\begin{equation}
\mathcal{L_\text{int}} = g_{s1} h^+_{3} \overline{\nu_\mu} \mu + g_{p1} h^+_{3} \overline{\nu_\mu} \gamma^5 \mu + h.c.
\end{equation}
In general, the contribution of a singly charged scalar to the muon anomaly can be expressed as,
\begin{equation}
\Delta a_\mu (h^+_{3}) \simeq \frac{1}{8\pi^2}\frac{m^2_\mu}{m^2_{h^+_{3}}} \int_0^1 dx \dfrac{g_{s1}^2 P_{s1}(x) + g_{p1}^2 P_{p1}(x)}{\epsilon^2 \lambda^2 (1-x) (1- \epsilon^{-2} x) +x}
\end{equation}
with $\epsilon \equiv \left(\frac{m_{\nu_\mu}}{m_{\mu}}\right),~ \lambda \equiv \left(\frac{m_\mu}{m_{h^+_{3}}}\right)$ and
\begin{center}
$ P_{s1}(x) = -x(1-x)(x + \epsilon) $\\
$ P_{p1}(x) = -x(1-x)(x - \epsilon) $
\end{center}
So, in this case the extra contribution is found to be,
\begin{equation}
\Delta a_\mu (h^+_{3}) \simeq -\frac{1}{4\pi^2}\frac{m^2_\mu}{m^2_{h^+_{3}}}\left[|g_{s1}^\mu|^2 \left(\frac{1}{12} \right) + |g_{p1}^\mu|^2 \left(\frac{1}{12}\right) \right];~~\text{with}~~ m_{h^+_3} \gg m_\mu , m_{\nu_\mu}
\end{equation}
\underline{\bf Fig.3(b): Contribution due to charged scalar, $h^+_4$ mediation; $\Delta a_\mu (h^+_{4})$} \\
 Similarly the interaction terms involving $h_4^+$ with scalar coupling ($g_{s2}$) and pseudo-scalar coupling ($g_{p2}$) are,
\begin{equation}
\mathcal{L_\text{int}} = g_{s2} h^+_{4} \overline{\nu_\mu} \mu + g_{p2} h^+_{4} \overline{\nu_\mu} \gamma^5 \mu + h.c.
\end{equation}
The expression for the contribution arising from this scalar to the muon anomaly can be written as,
\begin{equation}
\Delta a_\mu (h^+_{4}) \simeq -\frac{1}{4\pi^2}\frac{m^2_\mu}{m^2_{h^+_{4}}}\left[|g_{s2}^\mu|^2 \left(\frac{1}{12} \right) + |g_{p2}^\mu|^2 \left(\frac{1}{12} \right) \right];~~\text{with}~~ m_{h^+_4} \gg m_\mu , m_{\nu_\mu}
\end{equation}
The couplings for the above two cases can be found from Eq.\ref{charged_scalar_mbasis} and are given in Table \ref{charged_scalars_couplings}.\\
\begin{table}[h!]
\centering
\begin{tabular}{|c|c||c|c|}
\hline
  Interaction Vertex & $ g_{s1} = g_{p1}$ & Interaction Vertex & $ g_{s2} = g_{p2}$
           \\[3mm]
\hline \hline 
$\overline{\nu}_1 \mu h_3^+$   & $\frac{Y_{22}}{2}V_{\mu 1}^{\nu\nu \ast}$   
&$\overline{\nu}_1 \mu h_4^+$   & $-~\frac{\tilde{Y_{22}}}{2}V_{\mu 1}^{\nu\nu \ast}$   
          \\[2mm]
 \hline
 $\overline{\nu}_2 \mu h_3^+$   & $\frac{Y_{22}}{2}V_{\mu 2}^{\nu\nu \ast}$    
&$\overline{\nu}_2 \mu h_4^+$   & $-~\frac{\tilde{Y_{22}}}{2}V_{\mu 2}^{\nu\nu \ast}$   
          \\[2mm]
 \hline
 $\overline{\nu}_3 \mu h_3^+$   & $\frac{Y_{22}}{2}V_{\mu 3}^{\nu\nu \ast}$   
&$\overline{\nu}_3 \mu h_4^+$   & $-~\frac{\tilde{Y_{22}}}{2}V_{\mu 3}^{\nu\nu \ast}$   
          \\[2mm]
 \hline
 $\overline{S}_1 \mu h_3^+$   & $\frac{Y_{22}}{2}V_{\mu 1}^{\nu S \ast}$    
&$\overline{S}_1 \mu h_4^+$   & $-~\frac{\tilde{Y_{22}}}{2}V_{\mu 1}^{\nu S \ast}$   
          \\[2mm]
 \hline
 $\overline{S}_2 \mu h_3^+$   & $\frac{Y_{22}}{2}V_{\mu 2}^{\nu S \ast}$    
&$\overline{S}_2 \mu h_4^+$   & $-~\frac{\tilde{Y_{22}}}{2}V_{\mu 2}^{\nu S \ast}$   
          \\[2mm]
 \hline
 $\overline{S}_3 \mu h_3^+$   & $\frac{Y_{22}}{2}V_{\mu 3}^{\nu S \ast}$    
&$\overline{S}_3 \mu h_4^+$   & $-~\frac{\tilde{Y_{22}}}{2}V_{\mu 3}^{\nu S \ast}$   
          \\[2mm]
  \hline        
  $\overline{N}_1 \mu h_3^+$   & $\frac{Y_{22}}{2}V_{\mu 1}^{\nu N \ast}$    
&$\overline{N}_1 \mu h_4^+$   & $-~\frac{\tilde{Y_{22}}}{2}V_{\mu 1}^{\nu N \ast}$   
          \\[2mm]
 \hline
 $\overline{N}_2 \mu h_3^+$   & $\frac{Y_{22}}{2}V_{\mu 2}^{\nu N \ast}$    
&$\overline{N}_2 \mu h_4^+$   & $-~\frac{\tilde{Y_{22}}}{2}V_{\mu 2}^{\nu N \ast}$   
          \\[2mm]
 \hline
 $\overline{N}_3 \mu h_3^+$   & $\frac{Y_{22}}{2}V_{\mu 3}^{\nu N \ast}$    
&$\overline{N}_3 \mu h_4^+$   & $-~\frac{\tilde{Y_{22}}}{2}V_{\mu 3}^{\nu N \ast}$   
          \\[2mm]
 \hline
\end{tabular}
\caption{Relevant couplings associated with the Feynman diagrams involving $h_3^-, h_4^-$ given in Fig \ref{charged_scalar}.}
\label{charged_scalars_couplings}
\end{table}
The diagrams in Fig.\ref{neutral_CP_even_scalar} are mediated by CP-even neutral scalars $h_1^0$ and $h_2^0$. \\
\underline{\bf Fig.4(a): Contribution due to CP-even scalar, $h^0_1$ mediation; $\Delta a_\mu (h^0_{1})$} \\
 In general if extra electrically neutral scalar fields are present in a model, they induce a shift in the leptonic magnetic 
moments via the following interactions:
\begin{equation}
\mathcal{L_\text{int}} = g_{s3} h^0_{1} \overline{\mu} \mu + ig_{p3} h^0_{1} \overline{\mu} \gamma^5 \mu
\label{h_1}
\end{equation}
From Eq.\ref{h_1} one can see that scalar and pseudo-scalar couplings shift $(g-2)_\mu$ by
\begin{equation}
\Delta a_\mu (h^0_{1}) \simeq \frac{1}{4\pi^2}\frac{m^2_\mu}{m^2_{h^0_{1}}} \int_0^1 dx \dfrac{g_{s3}^2 P_{s3}(x) + g_{p3}^2 P_{p3}(x)}{(1-x) (1- \lambda^2 x) +\lambda^2 x}
\end{equation}
with $\lambda \equiv \left(\frac{m_\mu}{m_{h^0_{1}}}\right)$ and $ P_{s3}(x) = x^2(2-x)$, $ P_{p3}(x) = -x^3 $.

So, from here we have the extra contribution to the anomalous magnetic moment as,
\begin{equation}
\Delta a_\mu (h^0_{1}) \simeq \frac{1}{4\pi^2}\frac{m^2_\mu}{m^2_{h^0_{1}}}\left[|g_{s3}^\mu|^2 \left(-\frac{7}{12} - \text{log}\lambda \right) + |g_{p3}^\mu|^2 \left(\frac{11}{12} + \text{log}\lambda \right) \right];~~~\text{with}~~~ m_{h^0_1} \gg m_\mu.
\label{neutral_scalar_anomaly}
\end{equation}
The result in Eq.\ref{neutral_scalar_anomaly} is for general neutral scalars with scalar and pseudo-scalar couplings in the 
regime $m_{\text{Neutral Scalar}} \gg m_\mu$ . The contribution coming from pure scalar can be derived from Eq.\ref{neutral_scalar_anomaly} by setting 
the pseudo-scalar coupling ($g_p$) to zero and that from pseudo-scalar by setting the scalar coupling ($g_s$) to zero.
By comparing with Eq.\ref{scalar_mbasis} we have the couplings
$ g_{s3} = \tilde{Y}_{22},~g_{p3} = 0 $.\\
\underline{\bf Fig.4(b): Contribution due to CP-even scalar, $h^0_2$ mediation; $\Delta a_\mu (h^0_{2})$} \\
 For this diagram the interaction Lagrangian can be written as 
\begin{equation}
\mathcal{L_\text{int}} = g_{s4} h^0_{2} \overline{\mu} \mu + ig_{p4} h^0_{2} \overline{\mu} \gamma^5 \mu 
\label{h_2}
\end{equation}
Similar to the previous case its contribution to the anomalous magnetic moment can be written as,
\begin{equation}
\Delta a_\mu (h^0_{2}) \simeq \frac{1}{4\pi^2}\frac{m^2_\mu}{m^2_{h^0_{2}}}\left[|g_{s4}^\mu|^2 \left(-\frac{7}{12} - \text{log}\lambda \right) + |g_{p4}^\mu|^2 \left(\frac{11}{12}+ \text{log}\lambda \right) \right];~~~\text{with}~~~ m_{h^0_2} \gg m_\mu.
\end{equation}
 From comparison with Eq.\ref{scalar_mbasis} the couplings are $ g_{s4} = Y_{22},~g_{p4} = 0 $.\\
\\
\underline{\bf Fig.5(a): Contribution due to CP-odd scalar, $\phi^0_1$ mediation; $\Delta a_\mu (\phi^0_{1})$} \\
 In this case the interaction Lagrangian is given by,
\begin{equation}
\mathcal{L_\text{int}} = g_{s5} \phi^0_{1} \overline{\mu} \mu + ig_{p5} \phi^0_{1} \overline{\mu} \gamma^5 \mu
\end{equation}
As in the case 4(a), here we will have the extra contribution to the anomalous magnetic moment as,
\begin{equation}
\Delta a_\mu (\phi^0_{1}) \simeq \frac{1}{4\pi^2}\frac{m^2_\mu}{m^2_{\phi^0_{1}}}\left[|g_{s5}^\mu|^2 \left(-\frac{7}{12} - \text{log}\lambda \right) + |g_{p5}^\mu|^2 \left(\frac{11}{12} + \text{log}\lambda \right) \right];~~~\text{with}~~~m_{\phi^0_{1}} \gg m_\mu
\end{equation}
The couplings here are $g_{s5} = 0,~g_{p5} = - \tilde{Y}_{22} $.\\
\underline{\bf Fig.5(b): Contribution due to CP-odd scalar, $\phi^0_2$ mediation; $\Delta a_\mu (\phi^0_{2})$} \\
 For this interaction the Lagrangian can be written as,
\begin{equation}
\mathcal{L_\text{int}} = g_{s6} \phi^0_{2} \overline{\mu} \mu + ig_{p6} \phi^0_{2} \overline{\mu} \gamma^5 \mu
\end{equation}
and its contribution to $\Delta a_\mu$ is,
\begin{equation}
\Delta a_\mu (\phi^0_{2}) \simeq \frac{1}{4\pi^2}\frac{m^2_\mu}{m^2_{\phi^0_{2}}}\left[|g_{s6}^\mu|^2 \left(-\frac{7}{12} - \text{log}\lambda  \right) + |g_{p6}^\mu|^2 \left(\frac{11}{12} + \text{log}\lambda \right) \right];~~~\text{with}~~~m_{\phi^0_{2}} \gg m_\mu
\end{equation}
The couplings for this case are $g_{s6} = 0,~g_{p6} = Y_{22} $.\\
\begin{figure}[h]
\centering
\includegraphics[scale=0.7]{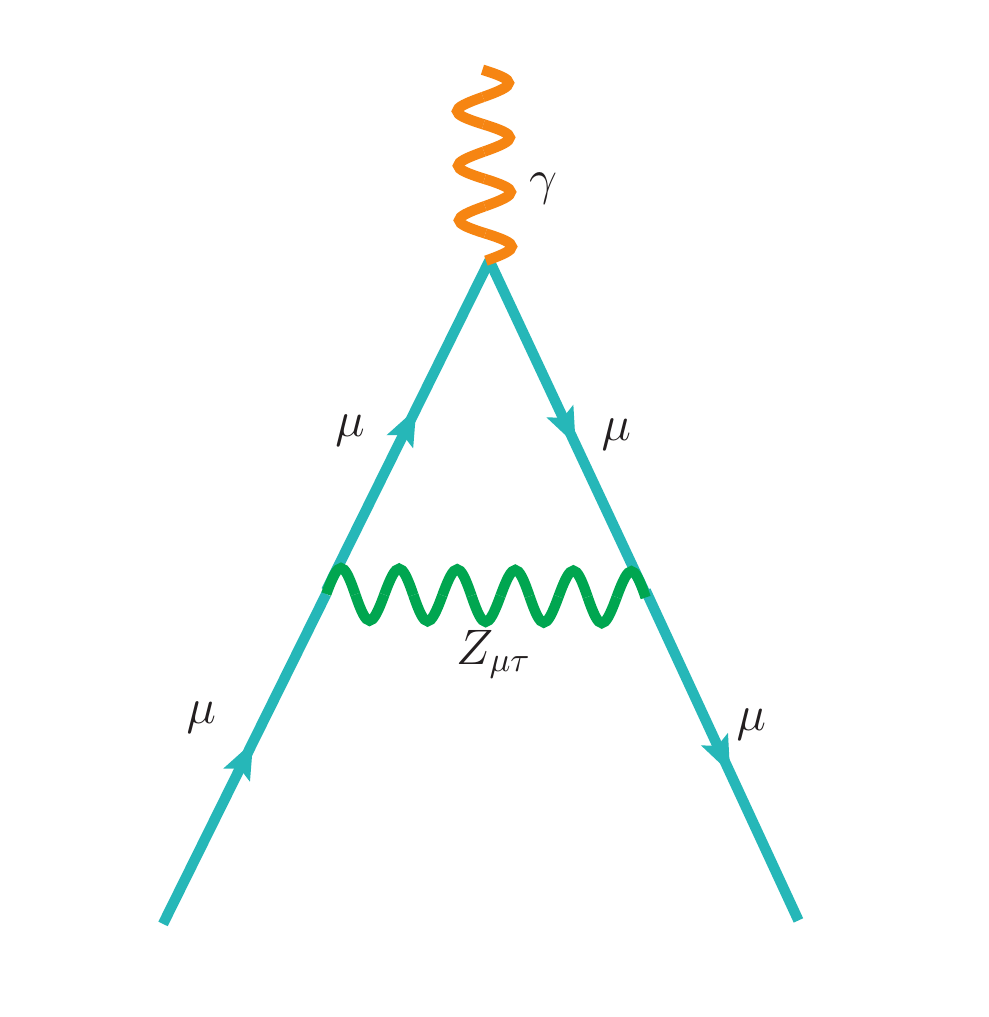}
\caption{Feynman diagram for the interaction of new light gauge boson $Z_{\mu\tau}$ with muons.}
\label{Zmt-mediation}
\end{figure}
\underline{\bf Fig.6: Contribution due to extra neutral gauge boson, $Z_{\mu\tau}$ mediation; $\Delta a_\mu (Z_{\mu \tau})$}\\
This diagram \ref{Zmt-mediation} comes from the interaction of the new gauge boson $Z_{\mu\tau}$ associated with $U(1)_{L_\mu-L_\tau}$ symmetry with muons.
We have the terms in the Lagrangian
\begin{equation}
\sum_{\alpha = e, \mu , \tau} \left[ \overline{\ell}_{\alpha L} \gamma^\mu D_\mu \ell_{\alpha L} + \overline{\ell}_{\alpha R} \gamma^\mu D_\mu \ell_{\alpha R} \right] 
 \end{equation}
 with covariant derivative $D_\beta = \partial_\beta + i g_{\mu \tau} q Z^{ \mu \tau}_{\beta} $, where $ g_{\mu \tau}$ is the gauge coupling of 
 $U(1)_{L_\mu - L_\tau}$ symmetry and $q$ is the corresponding $L_\mu-L_\tau $ charge ($q_{\mu,\nu_\mu} =1, q_{\tau,\nu_\tau}=-1$). By expanding this term explicitly 
 for $\mu$-family we will get $g_{\mu \tau} \overline{\mu} \gamma^\beta \mu Z^{ \mu \tau}_\beta $ and this term contributes to muon $(g-2)$ anomaly. \\
So, the interaction Lagrangian can be written as,
\begin{equation}
\mathcal{L_\text{int}} = g_{\mu \tau} Z_{\mu \tau} \overline{\mu} \gamma^\mu \mu
\end{equation}
Defining the parameter $\lambda \equiv \left(\frac{m_\mu}{m_{Z_{\mu \tau}}}\right)$, its contribution to the anomaly can be written as,
\begin{equation}
\Delta a_\mu (Z_{\mu \tau}) \simeq \frac{g^2_{\mu \tau}}{8\pi^2} \frac{m^2_\mu}{m^2_{Z_{\mu \tau}}} \int_0^1 dx \dfrac{2x^2(1-x)}{(1-x) (1- \lambda^2  x) +\lambda^2 x}
\end{equation}
After simplifying the integrations its contribution can be written as,
\begin{equation}
\Delta a_\mu (Z_{\mu \tau}) = \frac{g^2_{\mu \tau}}{12\pi^2}\frac{m^2_\mu}{m^2_{Z_{\mu \tau}}};~~~ \text{with}~~ \lambda \equiv \left(\frac{m_\mu}{m_{Z_{\mu \tau}}}\right)
\end{equation}

\section{Results and Discussion}
\label{sec:results}
Using the analytical expressions for different Feynman diagrams given in Sec \ref{sec:muon_anomaly_prediction}, we plot the dependence of $\Delta a_\mu$ on the masses of the various species. For the purpose of understanding the behaviour we retain a large range for each of the mass values, although as we see much of it is excluded by the collider data. The excluded regions are clearly marked out.
We see that the contribution of each of the class of diagrams independently could explain the entire anomaly,  however for several of the species the mass value that would have allowed this is already ruled out.

We use the data
$m_{W_L} = 80.4~\text{GeV},  m_Z = 91.2~\text{GeV}$, 
while for charge scalars we use the bounds \cite{Tanabashi:2018oca},
\begin{equation}
m_{h^+} > 181~\text{GeV}, m_{h^0,\phi^0} > 389~\text{GeV}.
\label{limits}
\end{equation}
 
For the standard results in the graphs the dashed green line represents the current bound 
on $\Delta a_\mu$. The red dashed lines represent the current $1\sigma$ bound on $\Delta a_\mu$. The values of these standard results \cite{Queiroz:2014zfa} are given below.
\begin{center}
$\Delta a_\mu(\text{Current Bound}) = (295 \pm 81)\times 10^{-11}$ \\
$\Delta a_\mu(1\sigma~ \text{Current Bound}) = 81\times 10^{-11} $
\end{center} 
It is useful to keep in mind the projected and 1$\sigma$ bounds on $\Delta a_\mu$ contribution which are $(295 \pm 34) \times 10^{-11}$ and $34 \times 10^{-11}$ respectively, since they may be soon reached, though we have not used them in our plots.

We include in our analysis the important possibility of asymmetric LRSM.
Usually in a left-right symmetric theory the $SU(2)_L$ and $SU(2)_R$ gauge couplings 
are equal, i.e. $g_L = g_R$, known as \textsl{symmetric LRSM} scenario. But, there is also the possibility that the Parity symmetry breaks at a higher scale than the $SU(2)_R$ gauge symmetry, in which case the left-handed and right-handed gauge couplings 
become unequal, i.e. $g_L \neq g_R$. Such a model is called \textsl{asymmetric LRSM}, which was first proposed in \cite{Chang:1983fu} and more about this 
can be found in \cite{Chang:1984uy,PhysRevD.31.1718,Sahu:2006pf,Bhattacharya:2006dn,Borah:2010zq,Majumdar:2020owj}. Hence, we have considered two 
different cases based on $g_L$ and $g_R$ for calculating the contributions of right-handed vector bosons $W_R$ and $Z_R$ to $\Delta a_\mu$.
\begin{center}
$\text{Case I :}~~ g_L = g_R = 0.653$ \\
$\text{Case II :}~~ g_L =  0.653, ~~~ g_R = 0.39$
\end{center}
where the latter case corresponds to Pati-Salam breaking scale of $10^6$~GeV, with grand unification in $SO(10)$ at $10^{17.2}$~GeV \cite{Majumdar:2020owj}.

With these representative values for gauge couplings we have numerically estimated and tabulated the upper bound on the muon anomaly contributions due to $W_R$ and $Z_R$ mediation channels (from their lower mass bound) for symmetric as well as asymmetric LRSM scenario in Table \ref{result_WRZR}.
\begin{table}[h!]
\centering
\begin{tabular}{|c|c|c|c|}
\hline
  Particles & Bounds on masses of mediators & $g_L = g_R~({\bf Case ~ I})$ &  $g_L\neq g_R~({\bf Case ~ II})$
           \\[3mm]
\hline \hline 
$\Delta a_\mu(W_R)$  & $ \geq$ 4.1 TeV \cite{Aad:2019hjw} & $\leq 1.45 \times 10^{-12}$   & $ \leq 0.55 \times 10^{-12}$
          \\[2mm]
 \hline
$\Delta a_\mu(Z_R)$ & $ \geq$ 4.9 TeV ( $g_L=g_R$) \& $\geq$ 9.0 TeV ($g_L \neq g_R$)  &  $\leq -~0.64 \times 10^{-12}$ & $\leq 0.10 \times 10^{-12}$
          \\[2mm]
  \hline
\end{tabular}
\caption{Estimated values of the individual contributions coming from $W_R$ and $Z_R$ in extended LRSM 
for the cases $g_L = g_R$ and $g_L \neq g_R$. Relation between $M_{W_R}$ and $M_{Z_R}$ in LRSM with Higgs doublets can be found in Ref.\cite{FileviezPerez:2016erl}, from which we have obtained the lower bound on $M_{Z_R}$ for our framework.} 
\label{result_WRZR}
\end{table}
\begin{figure}[h!]
\centering
\includegraphics[scale=0.6]{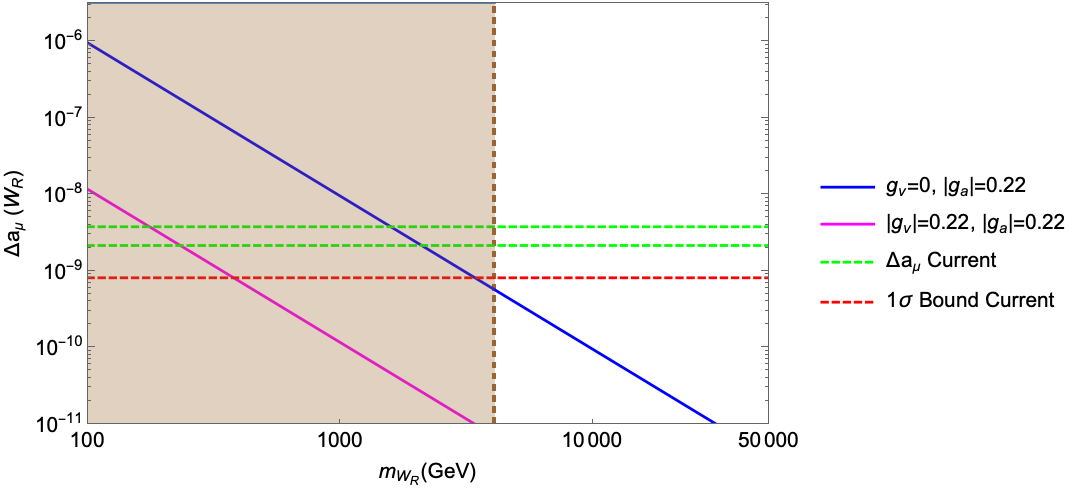}
\caption{Plot showing the contribution of charged vector boson $W_R$ to $\Delta a_\mu$ for the case $g_L=g_R$. The blue line represents 
the contribution of $W_R$ when purely axial vector-like coupling is considered and magenta line represents the contribution when both 
vector-like and axial vector-like couplings are considered non-zero. It shows with purely axial vector-like coupling $W_R$ with mass 
2 TeV can address the anomaly whereas the case with combination of both couplings can satisfy the muon anomaly bound for $m_{W_{R}} \sim 200$ GeV. Brown shaded region indicates the excluded mass range of $W_R$ from collider constraints. So both the muon anomaly contributions is ruled out from collider constraints on $m_{W_{R}}$.}
\label{fig:mWRsymm}
\end{figure}
\begin{figure}[h!]
\centering
\includegraphics[scale=0.6]{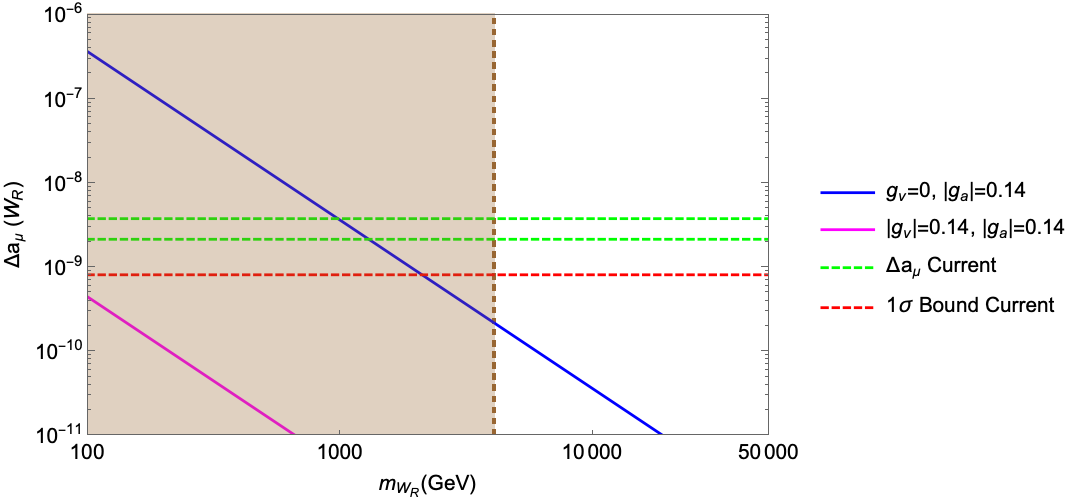}
\caption{Plot showing the contribution of charged vector boson $W_R$ to $\Delta a_\mu$ for the case $g_L \neq g_R$. The blue line represents 
the contribution of $W_R$ when purely axial vector-like coupling is considered and it is sensitive to current bound 
on $\Delta a_\mu$ when $W_R$ mass lies around 1 TeV. The magenta line represents the contribution when both 
vector-like and axial vector-like couplings are considered non-zero which is sensitive to the current bound on $\Delta a_\mu$ for mass range < 100 GeV but these two cases do not even satisfy the bound on $W_R$ mass (as they both appear within the brown shaded excluded mass range).}
\label{fig:mWRasymm}
\end{figure}

The contributions arising from charged gauge boson $W_R$ for the cases (i) $g_L = g_R$ (symmetric case), (ii) $g_L \neq g_R$ (asymmetric case) are presented in 
figures \ref{fig:mWRsymm} and \ref{fig:mWRasymm} respectively. 

(i) For the case $g_L = g_R$, if we consider purely axial-vector like coupling i.e. $|g_v| = 0$ and $|g_a| =0.22$ then the gauge boson $W_R$ with mass around 2 TeV
can address the whole anomaly. This is represented by the blue solid line in Fig.\ref{fig:mWRsymm}. whereas if we consider non-zero values for both couplings; $|g_v| = 0.22$ and $|g_a| =0.22 $, then the mass of $W_R$ lies 
around 200 GeV (magenta line). Thus both the cases fall in the excluded mass range of $W_R$ (brown shaded region) from collider bound.

(ii) Similarly for the case $g_L \neq g_R$, when purely axial-vector like coupling is considered i.e. $|g_v| = 0$ and $|g_a| =0.14$ then $W_R$ with mass around 1 TeV 
can explain the entire anomaly and the same is represented by blue line in Fig.\ref{fig:mWRasymm}.
But for $|g_v| = 0.14$ and $|g_a| =0.14 $ the mass of $W_R$ lies below 100 GeV (magenta line).
This implies that even though $W_R$ can explain the entire anomaly in both symmetric as well as asymmetric case, it is irrelevant for calculation since such a low mass 
for $W_R$ is ruled out by  collider experiments (brown shaded region) \cite{Aad:2019hjw}. 
Though from the experimental side, where $W_R$ interacts only with right handed neutrinos, i.e for $g_a = g_v$ the LEP bound on $\frac{g_{v}}{m_{W_R}}$ 
reads as $\frac{g_{v}}{m_{W_R}} < 4.8 \times 10^{-3}~\text{GeV}^{-1}$ \cite{Freitas:2014pua}. In our case $\frac{g_{v}}{m_{W_R}} \sim  5.1 \times 10^{-5}~\text{GeV}^{-1}$ 
which clearly satisfies the bound.

\begin{figure}[h]
\centering
\includegraphics[scale=0.6]{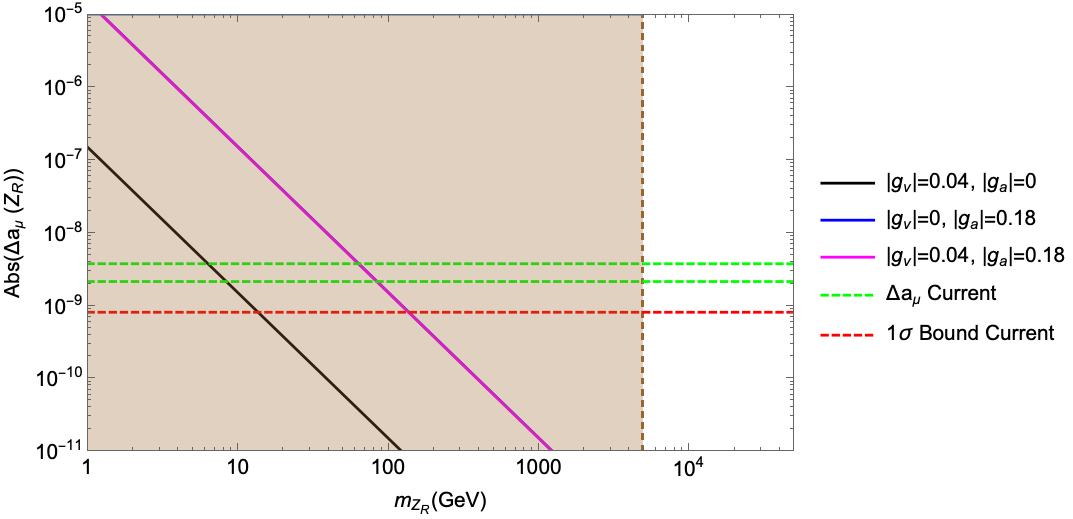}
\caption{Plot showing the contribution of neutral vector boson $Z_R$ to $\Delta a_\mu$ for the case $g_L=g_R$. The black line represents 
the contribution of $Z_R$ when purely vector-like coupling is considered and the magenta line represents the contribution when both 
vector-like and axial vector-like couplings are considered non-zero. The contribution with purely axial vector-like coupling is negative so we have plotted the absolute value of it here and it merges with magenta line. Even though the contribution from $Z_R$ is positive with purely vector-like coupling, all the cases fail to satisfy the current bound 
on $Z_R$ mass (as recent collider developments exclude the brown shaded mass range for $Z_R$).}
\label{fig:mZRsymm}
\end{figure}

\begin{figure}[h]
\centering
\includegraphics[scale=0.6]{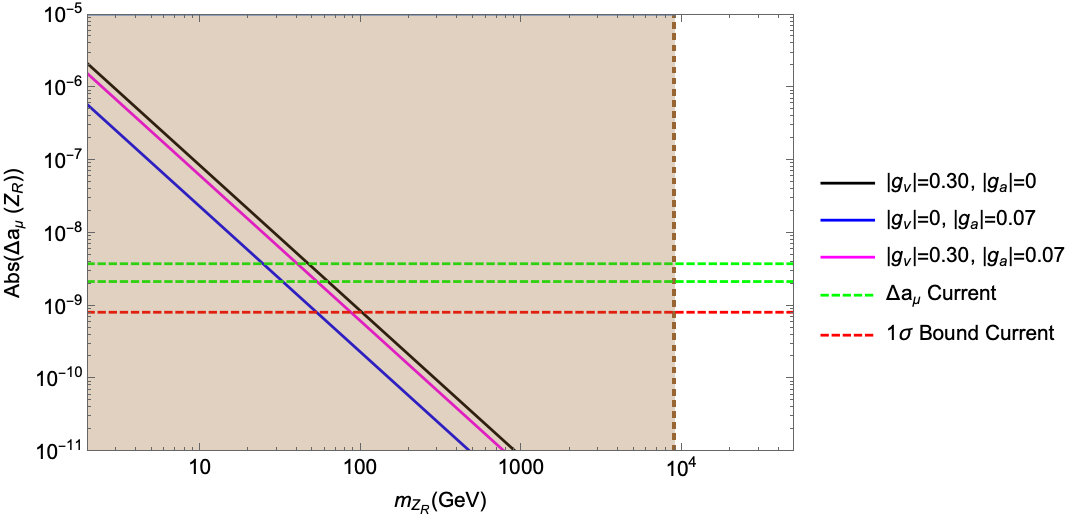}
\caption{Plot showing the contribution of neutral vector boson $Z_R$ to $\Delta a_\mu$ for the case $g_L \neq g_R$. The black line, blue line and magenta line 
represent the contributions of $Z_R$ when purely vector-like, purely axial vector-like and the combination of both couplings are considered non-zero 
respectively. Even though all contributions are positive and sensitive to the current bounds on $\Delta a_\mu$, none of the cases 
satisfy the bound on $Z_R$ mass since all the mass values which can satisfy the muon anomaly bounds are residing in the brown shaded excluded mass range.}
\label{fig:mZRasymm}
\end{figure}

Figures \ref{fig:mZRsymm} and \ref{fig:mZRasymm} show the contributions coming from the right-handed neutral vector boson $Z_R$ for the 
cases $g_L = g_R$ and $g_L \neq g_R$ respectively. For the case $g_L = g_R$, $Z_R$ gives negative contribution to $\Delta a_\mu$ and thus it is not 
relevant for our calculation, but for comparison perspective we have plotted the absolute value of the contributions 
vs $Z_R$ mass in Log-Log plots. We have shown the excluded mass range for $Z_R$ due to collider constraints as the brown shaded region in both these plots.

(i) For the case $g_L = g_R$, 
$Z_R$ contributes positively and could have addressed the anomaly with $M_{Z_R} \sim 10$ GeV 
when purely vector-like contribution is considered (black line in Fig.\ref{fig:mZRsymm}), but which lies deep in the excluded region. The other two contributions i.e. purely axial-vector 
like (blue line) and combination of both couplings (magenta line) give negative contributions.
 
(ii) For the case $g_L \neq g_R$, $Z_R$ contributes positively for all the choices on couplings i.e. purely vector-like (black line in Fig.\ref{fig:mZRasymm}), purely axial-vector-like 
(blue line) as well as combination of both (magenta line) and can explain the anomaly but with $M_{Z_R} \sim 20-50$ GeV, far below the collider bounds. This accords with ref.\cite{Freitas:2014pua} which argues that a 95\% C.L upper bound from LEP measurements applies for $g_{v} = g_{a}$ and $m_{Z_R} > \sqrt{s}$ that puts $\frac{g_{v}}{m_{Z_R}} < 2.2 \times 10^{-4}~\text{GeV}^{-1}$ and thus discards the idea of a single $Z_R$ boson explaining the anomaly. Some more bounds are given in ref. \cite{Freitas:2014pua, Beringer:1900zz}.

\begin{figure}[h]
\centering
\includegraphics[scale=0.6]{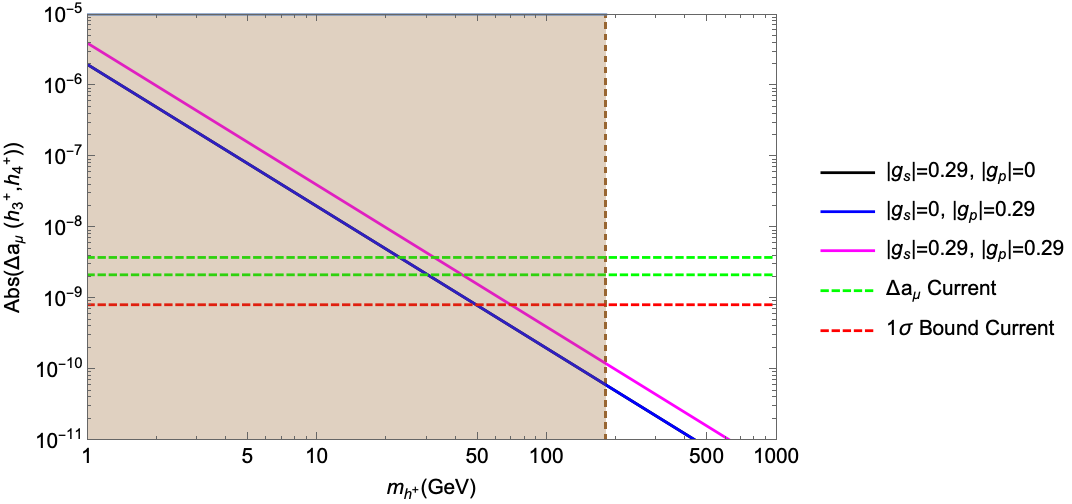}
\caption{Plot showing the contributions of singly charged scalars $h_3^+, h_4^+$ to $\Delta a_\mu$ for three different choices of couplings; 
 purely scalar, purely pseudo-scalar and combination of both. All the contributions are negative and thus are plotted their absolute values in log-log plot. 
 The contributions coming from purely scalar and purely pseudo-scalar couplings are super-imposed and represented by the blue line.
 The magenta line represents the contribution from the combination of both couplings. Also the mass range for charged scalars $<$ 181 GeV is ruled out from collider studies (brown shaded region).}
\label{fig:mH+}
\end{figure}

Figure \ref{fig:mH+} shows the contributions coming from the charged scalars $h_3^+, h_4^+$ for three different choices of the couplings; 
$|g_s| = 0.29$ and $|g_p| =0 $ (purely scalar), $|g_s| = 0$ and $|g_p| =0.29 $ (purely pseudo-scalar) and $|g_s| = 0.29$ and $|g_p| =0.29 $ (combination of both). 
We have already discussed in Sec \ref{sec:muon_anomaly_prediction} that $h_3^+, h_4^+$ contribute negatively to $\Delta a_\mu$, and thus we have plotted 
the absolute values of these contributions in Log-Log plot. Here the black and blue lines representing purely scalar and purely pseudo-scalar couplings 
which coincide together. Magenta line represents the contribution coming from the charged scalar sector when 
we consider both the scalar as well as pseudo-scalar couplings non-zero. 
The plot shows that the masses of the charged scalars lie around $\mathcal{O}$(50) GeV which cannot satisfy the collider bounds on masses given in relation \ref{limits} (brown shaded region of the plot shows the excluded range). 
Also from the results it can be concluded that singly charged scalars are not good candidates for explaining
muon $(g-2)$ anomaly since they give negative and suppressed contribution.

\begin{figure}[h]
\centering
\includegraphics[scale=0.5]{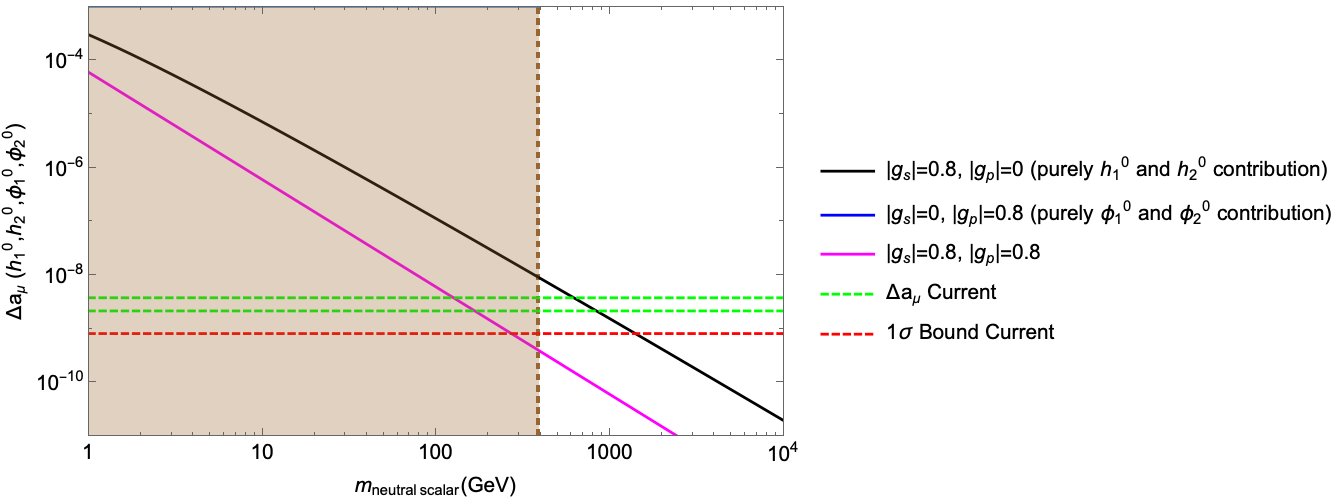}
\caption{Plot showing the contributions of CP-even and CP-odd neutral scalars to $\Delta a_\mu$. 
 The black line represents purely scalar contribution coming from $h_1^0, h_2^0$, while 
 the magenta line represents the combination of both scalar as well as pseudoscalar couplings. For purely pseudo-scalar coupling the contribution becomes negative (we have not shown it in this plot). Brown shaded region indicates the excluded mass range for neutral scalars.}
\label{fig:mH0}
\end{figure}
Figure \ref{fig:mH0} shows the contributions coming from all the neutral scalars present in our model (all of them arising from bidoublet $\Phi$), namely $h_1^0, h_2^0, \phi_1^0, \phi_2^0$. 
We have mentioned earlier that the contribution to muon anomaly coming from either pure scalar or pure pseudo-scalar or both can be easily derived from their couplings. 
In this case we can see that the neutral CP-even scalars $h_1^0$ and $h_2^0$ with mass around 500 GeV can explain the entire anomaly if we consider 
pure scalar couplings i.e., $g_s=0.8$ and $g_p=0$ for them (represented by black line).
If we consider purely pseudo-scalar coupling; i.e. $g_s$ = 0 and $g_p$ = 0.8 
then the contribution becomes negative (we have not plotted this contribution in figure \ref{fig:mH0}). However if we take non-zero values for both the scalar and pseudo-scalar couplings, 
i.e. $g_s$ = 0.8 and $g_p$ = 0.8 (represented by the magenta line), then a neutral scalar with 150 GeV mass can address the anomaly since it is sensitive to  
the bounds on $\Delta a_\mu$. But only the CP-even scalars can satisfy both muon anomaly as well as allowed mass range constraints for neutral scalars in one go (excluded mass range in the figure is indicated by brown shaded region). However considering massive CP even scalars with mass around $\mathcal{O}(10)$ TeV or higher, though saturate the allowed mass range constraint, fail to satisfy the current bound on muon anomaly. This can also be easily inferred from the plot. It is to be noted that neutral scalars are constrained by LEP searches for four-lepton contact interactions which requires $ \frac{g}{M_\phi} < 2.5 \times 10^{-4}~\text{GeV}^{-1}$ for $M_\phi > \sqrt{s}$ \cite{Freitas:2014pua}. For our case $ \frac{g}{M_\phi} = 1.6 \times 10^{-4}~\text{GeV}^{-1}$ which clearly satisfies the LEP search bound.

\begin{figure}[h]
\centering
\includegraphics[scale=0.65]{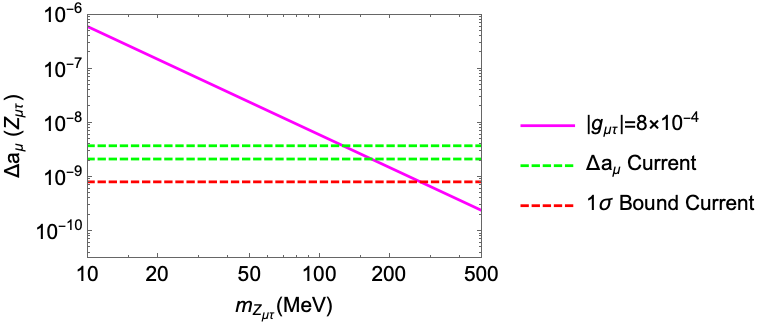}
\caption{Plot showing the contribution coming from new light gauge boson $Z_{\mu \tau}$ $vs$ mass of $Z_{\mu \tau}$. The magenta line shows the contribution of $Z_{\mu \tau}$ 
with coupling $g_{\mu \tau} =8 \times 10^{-4}$ can address the anomaly with $Z_{\mu \tau}$ mass lying around 150 MeV. }
\label{fig:mZmt}
\end{figure}
Figure \ref{fig:mZmt} shows the contribution coming from the new neutral vector boson $Z_{\mu \tau}$ in our model that comes from the $U(1)_{L_\mu -L_\tau}$ extension of LRSM. 
The plot shows that for coupling $g_{\mu \tau} =8 \times 10^{-4}$, the neutral vector boson $Z_{\mu \tau}$ having mass nearly 150 MeV can address the entire anomaly (magenta line).
The coupling strength ($g_{\mu \tau}$) of this vector boson is strongly constrained to be less than $\simeq 10^{-3}$ from 
the measurement of neutrino trident cross section by experiments like CHARM-II \cite{GEIREGAT1990271} and CCFR \cite{PhysRevLett.66.3117} while a mass of $\mathcal{O}$(100 MeV) is 
allowed, and both of these are satisfied in our case. 

\begin{figure}[h]
\centering
\includegraphics[scale=0.65]{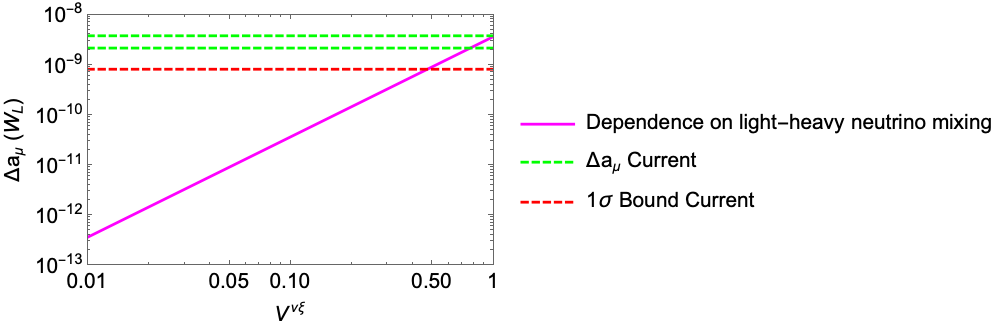} 
\caption{Plot showing the variation of $\Delta a_\mu$ coming from purely left-handed currents via $W_L$ mediation $\mbox{vs.}$ the light-heavy mixing parameter $V^{\nu \xi}$.}
\label{fig:mWL}
\end{figure}

In general, the individual contribution to muon anomaly arising from a mediating particle is related to its mass by the relation, 
\begin{equation}
\Delta a_\mu \propto \frac{1}{m_{\text{mediator}}^2}
\end{equation}
Thus all the contributions to $\Delta a_\mu$ become negligible in our model except those of $W_L$ and $Z_{\mu \tau}$, if we consider heavy mass for the 
particles which are allowed by the collider experiments.  The significant contribution comes from $W_L$ if we consider large light-heavy neutrino mixing 
which is facilitated by inverse seesaw mechanism in the model. 
However, the light-heavy neutrino mixing parameter $V^{\nu \xi}$ can be constrained by other sectors like non-unitarity effects in experiments looking for lepton flavour violation and NSI effects at neutrino factory as discussed in section\,\ref{subsec:NSI}.

Figure \ref{fig:mWL} shows how the contribution of $W_L$ to $\Delta a_{\mu}$ varies with different mixing values. Here we vary this mixing from $10^{-2}$ 
to 1. The magenta line represents the dependence of $\Delta a_\mu$ on light-heavy mixing and we find that $V^{\nu \xi}$
should be of $\mathcal{O}(0.3-1)$ in order to satisfy current bound on $\Delta a_\mu$. 
A complete analysis of the results from the plots as well as the tables shows that significant contributions 
to $\Delta a_\mu$ come from from $W_L$ and $Z_{\mu\tau}$ in the model whereas all other contributions are either negative, suppressed or ruled out by collider limits. However lighter neutral CP even scalars with mass around $0.5-2$ TeV can also be a good candidate to satisfy the entire current muon anomaly bound individually.
Were the light-heavy neutrino mixing to be large in the inverse seesaw framework, $W_L$ contribution could have accounted for the entire muon anomaly \cite{Tanabashi:2018oca, Dev:2020drf} individually when the 
mixing ($V^{\nu \xi}$) is $\mathcal{O}(1)$.  
However the present bounds on the $| \eta|$ parameters of Sec. \ref{subsec:NSI} allow $\lesssim 0.3$ as the optimistic value of this parameter. 
If these bounds are confirmed this contribution is still capable of explaining approximately $10\%$ of the anomaly as can be seen from Fig. \ref{fig:mWL}.
Also for light extra neutral gauge boson contribution, we can easily infer that $m_{Z_{\mu\tau}} \sim 150$ MeV and coupling $g_{\mu\tau} \sim 8 \times 10^{-4}$ can ameliorate the entire anomaly. Equally importantly we have thus established that in case natural values of the parameters of any one contribution are insufficient, the three together (i.e., contributions coming from CP even scalars $h_1^0, h_2^0$, singly-charged gauge boson $W_L$ and new light neutral gauge boson $Z_{\mu\tau}$ mediation channels) have the potential to explain the entire anomaly within our ELISS scenario.

\section{Conclusion}
\label{sec:conclusion}
We have studied the $U(1)_{L_\mu-L_\tau}$ extension of left-right symmetric model which can explain non-zero neutrino mass, mixing and muon anomalous 
magnetic moment within a single framework. Neutrino mass is generated in the model through inverse seesaw mechanism that allows large light-heavy neutrino mixing. 
We have discussed how the choice of scalars in various LRSM-SM symmetry breaking 
chains affect the generation of 
neutrino mass. We have calculated the individual contributions due to all the vector bosons and scalars present in the model to muon anomaly and found out that vector boson $W_L$ with $\mathcal{O}(1)$ light-heavy neutrino mixing, 
the new light neutral vector boson $Z_{\mu\tau}$ as well as low-massive CP-even scalars are good candidates for explaining the entire anomaly. Although 
$W_L$'s interaction with heavy right-handed neutrino, facilitated by inverse seesaw mechanism, becomes one of the 
significant contributions to the anomaly as this can account for upto $10\%$ of the entire anomaly if one considers the constraints from NSI.  
Another major contribution comes from the new gauge boson $Z_{\mu\tau}$ which can explain the whole anomaly with mass $150$ MeV and coupling $g_{\mu \tau} =8 \times 10^{-4}$. 
The contributions coming from $Z_R$ for different choices of couplings are negative whereas those of $W_R$ are positive but invalid since it does not satisfy 
the allowed mass range. We have also briefly presented the comparative study of the effects between symmetric and asymmetric LRSM scenarios to muon anomaly estimation. Similarly the contributions arising from the charged as well as CP-odd neutral scalars are either suppressed or negative whereas CP-even neutral scalars can satisfy the entire muon anomaly bound for mass range $\sim$ 500 GeV, but considering massive scalars with mass $\sim \mathcal{O}(10)$ TeV or higher, we will get negligible contribution to muon anomaly due to neutral scalar mediation.
We have also shown in plots how the contribution of each particle to $\Delta a_\mu$ varies with the mass of that particle for different choices of couplings.
Overall we have found that inverse seesaw mechanism influences the results on muon anomaly to a large extent while explaining neutrino mass and 
mixing simultaneously in the model.

\section{Acknowledgement}
SS is thankful to UGC for fellowship grant to support her research work. 
\bibliographystyle{JHEP}
\bibliography{Muon_Anomaly}
\end{document}